\documentclass[12pt]{article} 
\usepackage{jheppub}

\usepackage{amsmath,amsfonts,amsbsy,amssymb,array,accents,dsfont}
\usepackage{enumerate}
\usepackage[shortlabels]{enumitem}
\usepackage{graphicx} 
\usepackage{subcaption} 
\usepackage{upgreek}
\usepackage{bm}
\usepackage{slashed}

\let\includefigures=\iftrue
\let\useblackboard==\iftrue
\newfam\black

\usepackage[dvipsames]{xcolor}

\definecolor{myblue}{RGB}{85,130,255}
\definecolor{myred}{RGB}{200, 45, 40}
\newcommand{\myBlue}{\color{myblue}}
\newcommand{\myRed}{\color{myred}}

\PassOptionsToPackage{ 
     colorlinks=true,
     linkcolor=myBlue,
     urlcolor=blue,
     filecolor=black,
     citecolor=myRed,
}{hyperref}

\usepackage{xparse}
\usepackage{xspace}

\NewDocumentCommand\eqn{om}{%
  \IfNoValueTF{#1}
     {\[ #2 \]}
     {\begin{equation}\label{#1} #2  \end{equation} \expandafter\newcommand\csname #1\endcsname{\eqref{#1}\xspace}\ignorespaces}
}
\NewDocumentCommand\eqna{om}{%
  \IfNoValueTF{#1}
    {\begin{align*} #2 \end{align*}}
    {\begin{equation}\label{#1}\begin{split} #2  \end{split}\end{equation} \expandafter\def\csname #1\endcsname{\eqref{#1}\xspace}\ignorespaces}
}

\newcommand{\rcite}{\cite}

\def\jb{{\,\pmb {\rm j}}}

\def\gammab{{\pmb\gamma}}

\def\epsilonb{{\boldsymbol\epsilon}}

\def\pib{{\boldsymbol\pi}} 
\def\sigmab{{\boldsymbol\sigma}} 
\def\taub{{\boldsymbol\tau}}

\def\gosc{\ensuremath{\mathfrak{g}}}
\def\g{{\mathsf g}}

\def\sigone{{\genfrac(){0pt}{1}{0\;1}{1\;0}}}

\def\sl{\text{sl}}
\def\su{\text{su}}

\def\sltwo{\ensuremath{SL(2,\bR)}}
\def\sltwoc{\ensuremath{SL(2,\bC)}}

\def\sutwo{{SU(2)}}

\def\uone{U(1)}

\def\sqsphere{{\bS^3_\flat}}

\def\tight#1{\! #1 \!}  

\def\({\left(}
\def\){\right)}
\def\[{\left[}
\def\]{\right]}

\def\ie{{i.e.}}
\def\eg{{e.g.}}
\def\cf{{c.f.}}
\def\etc{{etc}}

\def\ST{{\sst\rm ST}}

\def\eff{{\rm eff}}

\def\max{{\rm max}}
\def\min{{\rm min}}

\def\lstr{\ell_{\textit{s}}}

\def\gstr{g_{\textit s}^{\;}}
\def\gstrsq	{g_{\textit s}^{2}}

\def\th{{\rm th}}

\def\ntil{{n-2}}
\def\ktil{{k+2}}
\def\etatil{\alpha}

\def\sfp{{\mathsf p}}
\def\sfq{{\mathsf q}}

\DeclareMathSymbol{\medhatsym}{\mathord}{largesymbols}{"62} 
\newcommand\lowermedhatsym{
  \text{\smash{\raisebox{-1.28ex}{%
    $\medhatsym$}}}}
\newcommand\medhat[1]{
  \mathchoice
    {\accentset{\displaystyle\lowermedhatsym}{#1}}
    {\accentset{\textstyle\lowermedhatsym}{#1}}
    {\accentset{\scriptstyle\lowermedhatsym}{#1}}
    {\accentset{\scriptscriptstyle\lowermedhatsym}{#1}}
}

\DeclareMathSymbol{\medtildesym}{\mathord}{largesymbols}{"65}

\makeatletter
\newcommand*\rel@kern[1]{\kern#1\dimexpr\macc@kerna}
\newcommand*\widebar[1]{%
  \begingroup
  \def\mathaccent##1##2{%
    \rel@kern{0.8}%
    \overline{\rel@kern{-0.8}\macc@nucleus\rel@kern{0.2}}%
    \rel@kern{-0.2}%
  }%
  \macc@depth\@ne
  \let\math@bgroup\@empty \let\math@egroup\macc@set@skewchar
  \mathsurround\z@ \frozen@everymath{\mathgroup\macc@group\relax}%
  \macc@set@skewchar\relax
  \let\mathaccentV\macc@nested@a
  \macc@nested@a\relax111{#1}%
  \endgroup
}
\makeatother

\def\Phihat{\medhat\Phi}

\def\half{\frac12}
\def\coeff#1#2{{\textstyle \frac{#1}{#2}}}
\def\hf{\coeff12}
\def\tr{{\rm Tr}}

\def\One{{\hbox{1\kern-1mm l}}}

\def\ch{{\rm ch}}
\def\sh{{\rm sh}}

\def\barray{\begin{array}}
\def\earray{\end{array}}
\def\be{\begin{equation}}
\def\ee{\end{equation}}
\def\bea{\begin{eqnarray}}
\def\eea{\end{eqnarray}}
\def\bal{\begin{align}}
\def\eal{\end{align}}
\def\nn{\nonumber}

\newcommand{\bC}{{\mathbb C}}

\newcommand{\bH}{{\mathbb H}}

\newcommand{\bN}{{\mathbb N}}

\newcommand{\bR}{{\mathbb R}}
\newcommand{\bS}{{\mathbb S}}
\newcommand{\bT}{{\mathbb T}}

\newcommand{\bZ}{{\mathbb Z}}

\definecolor{cardinal}{rgb}{0.6,0,0}
\definecolor{darkgreen}{rgb}{0,0.4,0}
\definecolor{green}{rgb}{0,0.4,0}
\definecolor{golden}{rgb}{0.92, 0.7, 0}
\definecolor{midnight}{rgb}{0, 0, 0.5}
\definecolor{darkblue}{rgb}{0, 0, 0.7}

\numberwithin{equation}{section}

\mathchardef\mhyphen="2D

\def\cA{\mathcal {A}}  \def\cC{\mathcal {C}}
\def\cD{\mathcal {D}} \def\cE{\mathcal {E}} 
\def\cG{\mathcal {G}}  \def\cI{\mathcal {I}}
  
\def\cM{\mathcal {M}} \def\cN{\mathcal {N}} \def\cO{\mathcal {O}}
  \def\cR{\mathcal {R}}
\def\cS{\mathcal {S}}  
\def\cV{\mathcal {V}}

\def\one{{\hbox{\kern+.5mm 1\kern-.8mm l}}}
\def\zero{{\hbox{0\kern-1.5mm 0}}}

\def\id{\textrm{id}}

\newcommand{\bl}[1]{{\myBlue{\pmb {#1}}}}
\newcommand{\rd}[1]{{\myRed{\pmb {#1}}}}

\def\id{{1 \kern-.28em {\rm l}}}

\def\journal#1&#2(#3){\unskip, \sl #1\ \bf #2 \rm(19#3) }
\def\andjournal#1&#2(#3){\sl #1~\bf #2 \rm (19#3) }

\def\ie{{\it i.e.}}
\def\eg{{\it e.g.}}
\def\cf{{\it c.f.}}

\def\etc{{\it etc}}

\def\sst{\scriptscriptstyle}

\def\coeff#1#2{{\textstyle{\frac{#1}{ #2}}}}
\def\half{\frac12}
\def\hf{{\textstyle\half}}

\def\One{{1\hskip -3pt {\rm l}}}

\catcode`\@=11
\def\slash#1{\mathord{\mathpalette\c@ncel{#1}}}
\def\underrel#1\over#2{\mathrel{\mathop{\kern\z@#1}\limits_{#2}}}

\catcode`\@=12

\def\exp{{\rm exp}}

\def\ie{{\it i.e.}}
\def\eg{{\it e.g.}}

\def\mbar{{\bar m}}

\def\LG{{\sst\bf LG}}

\title{
{
A Defect in \texorpdfstring{$\bf AdS_3/CFT_2$}{} Duality
}}

\author{
Emil J. Martinec
}

\affiliation{
\vskip 0.01cm
Kadanoff Center for Theoretical Physics and Enrico Fermi Institute\\ 
University of Chicago\\ 
5640 S. Ellis Ave.\\
Chicago IL 60637\\ 
}

\emailAdd{%
e-martinec@uchicago.edu}

\abstract{%
$AdS_3$ string theory in the stringy regime $k=(R_{AdS}/\lstr)^2 < 1$ provides a laboratory for the study of holography in which both sides of AdS/CFT duality are under fairly good control.  
Worldsheet string theory is solvable, and for closed strings the dual spacetime CFT is a deformation of a symmetric product orbifold.
Here we extend this construction to include open strings by adding a probe D-string, described semiclassically by an $AdS_2$ D-brane in $AdS_3$.  The dual defect or boundary conformal field theory (BCFT) is again a deformed symmetric product, which now describes the Fock space of long open and closed strings near the AdS boundary, with a boundary deformation implementing the open/closed transition in addition to the symmetric product $\bZ_2$ twist deformation that implements closed string joining/splitting.  The construction thus provides an explicit example of an $AdS_3/BCFT_2$ duality. 
}

\begin{document}
\hypersetup{pageanchor=false}
\begin{titlepage}
\maketitle
\thispagestyle{empty}
\end{titlepage}
\hypersetup{pageanchor=true}
\pagenumbering{arabic}

\thispagestyle{empty}

\vskip 1cm
\hrule


\section{Introduction} 
\label{sec:intro}



The $AdS_3/CFT_2$ correspondence is an especially fruitful laboratory for the exploration of gauge/gravity duality.  On the CFT side, conformal symmetry is especially powerful in two dimensions~\rcite{DiFrancesco:1997nk}, and on the AdS side one has available the tools of worldsheet string theory~\rcite{Giveon:1998ns,Kutasov:1999xu,Maldacena:2000hw,Maldacena:2001km} to explore beyond the supergravity limit.

One of the key achievements of gauge/gravity duality is the insight it has provided into the quantum nature of black holes.  However, due to the difficulty of decoding the gravitational description from that of the gauge theory, an understanding of black hole structure at the horizon scale, and the nature (or even the existence) of the block hole interior, remain elusive.

For this reason it can be useful to study the threshold of black hole formation from nearby regimes.  One such regime is that of BPS and near-BPS configurations (see for instance~\rcite{Mathur:2005zp,Bena:2013dka,Shigemori:2020yuo} for reviews). There is an enormous number of such configurations even if the amount is insufficient to generate a ``geometric'' entropy equivalent to a effective area $A_\eff$ much larger than the Planck or string scale~\rcite{Shigemori:2019orj}.  In addition, by paying sufficient attention to stringy effects one can get a glimpse of the degrees of freedom responsible for black hole entropy from the bulk perspective~\rcite{Martinec:2019wzw,Martinec:2021vpk}.

Another nearby regime is that string scale curvature $k=(R_{AdS}/\lstr)^2<1$, where the high energy density of states 
\eqn[cardy]{
S \sim 2\pi \sqrt{(k p -h_{\rm min}/4)L_0} + 2\pi \sqrt{(k p -h_{\rm min}/4)\bar L_0}
~~,~~~~
h_{\rm min} \approx \frac{p(k-1)^2}{4k}
}
is dominated by strings rather than black holes~\rcite{Giveon:2005mi}, 
(here $p$ is the fundamental string charge of the background)
but approaches the black hole density of states as $k\to 1^-$.
This stringy regime was investigated in detail in~\rcite{Balthazar:2021xeh}.  Worldsheet string theory realizes the regime $k<1$ via a noncritical background
\eqn[noncrit]{
\sltwo_k \times \sutwo_n^\flat
~~,~~~~
k=\frac{n}{n+1} ~,
}
where the subscripts denote the levels of the corresponding current algebras, and the ``flat'' superscript indicates that the $\sutwo$ factor must be deformed by a particular  $\int\! J^3_\su\bar J^3_\su$ interaction in order to preserve spacetime supersymmetry.  This deformation ``squashes'' the round three-sphere geometry of $\sutwo$, hence we refer to the background as ``squashed $\bS^3$''.

The CFT dual to this background takes the form of a symmetric product orbifold
\eqn[symprod]{
\big(\cM\big)^p/S_p
~~,~~~~
\cM = \bR_\phi \times \bS^3_\flat ~,
}
which is deformed by a marginal $\bZ_2$ twist operator.  Here $\bR_\phi$ is parametrized by a free field $\phi$ with a linear dilaton having slope
\eqn[Qlong]{
Q_\ell = \sqrt\frac{2}{n(n+1)}~,
}
and $\bS^3_\flat$ is the same squashed three-sphere theory as on the worldsheet.  
Thus the spacetime CFT has central charge
\eqn[cST]{
c_{\ST} = 6kp ~.
}
The block CFT $\cM$ describes the degrees of freedom of transverse to a long fundamental string of unit winding in the geometry~\eqref{noncrit}, in particular $\phi$ parametrizes the radial direction in $AdS_3$.  This feature is quite useful for understanding the duality~-- usually the radial location of excitations is rather indirectly and obscurely specified in the CFT description.

The symmetric product structure is similar to that of the spacetime CFT dual to the critical dimension background $AdS_3\times\bS^3\times \bT^4$ at $k=1$, except that the building block of the symmetric product there is the $\bT^4$ of the compactification wrapped by the fivebranes, while here the block theory~\eqref{symprod} describes directions transverse to the background fivebranes.  While there are highly twisted sectors in both cases which have been associated to a black hole spectrum~\rcite{Maldacena:1996ds}, for $k<1$ these sectors describe Hagedorn strings vibrating in the ambient spacetime rather than black holes composed of little strings trapped inside the fivebranes~\rcite{Balthazar:2021xeh,Martinec:2021vpk}.  Thus it is of interest to investigate the similarities and differences between the two and how they are manifested in the properties of the spacetime CFT.

Another construction that has been associated to black hole microstates is the ``end-of-the-world'' brane%
~\rcite{
Almheiri:2019hni,Sully:2020pza,Chen:2020uac,Chen:2020hmv}, whose spacetime CFT dual is supposed to be a boundary CFT (BCFT).  More generally, boundary and defect CFT's have proven a useful tool in the exploration of gauge/gravity duality%
~\rcite{Bachas:2001vj,Chiodaroli:2010mv,Chiodaroli:2011nr,Chiodaroli:2011fn,Takayanagi:2011zk,Fujita:2011fp,Jensen:2013lxa,Erdmenger:2014xya,Miyaji:2014mca,Gutperle:2015hcv,Bachas:2020yxv,Miyaji:2021ktr,Geng:2021iyq}

In this work, we investigate BCFT duals to $AdS_3$ string theory with $k<1$.  We will mainly consider an $AdS_2$ brane worldvolume in $AdS_3$~\rcite{Bachas:2000fr,Lee:2001xe,Lee:2001gh,Ponsot:2001gt,Giveon:2001uq,Karch:2000gx,Israel:2005ek,Hosomichi:2006pz}.  
We begin in section~\ref{sec:classical} with an analysis of classical solutions for open strings on an $AdS_2$ brane, following~\rcite{Lee:2001xe}.  While we are eventually interested in the stringy regime $k<1$ and the classical limit is the opposite limit $k\gg1$, the classical dynamics characterizes the various classes of string states, and thus serves as a useful lead-in to an overview of the quantum theory in section~\ref{sec:quantumcomments}.  

The spacetime CFT dual to this open string worldsheet theory once again is a symmetric product that describes the Fock space of long open and closed strings.  There are two sorts of primitive long open strings, asymptotically stretching halfway around either side of $AdS_2\subset AdS_3$, and so the symmetric product takes the form
\eqn[opensymprod]{
\big[\big(\cM^+\big)^{p_+}\tight\times \big(\cM^-\big)^{p_-}\big]/\big(S_{p_+}\tight\times S_{p_-}\big)  ~,
}
with $p_+$ being the number of long strings stretching around one side and $p_-$ the number stretching around the other side.
These open strings interact in the bulk of their worldvolume by the same sort of marginal $\bZ_2$ twist deformation from~\rcite{Balthazar:2021xeh} that describes closed string interactions.  In addition, there are open-closed transitions that sew together the primitive open strings into open and closed strings of higher winding.  The description of these longer strings in the above symmetric product is the subject of section~\ref{sec:OpenSymProd}, and the boundary operator that implements string endpoint creation/annihilation is described in section~\ref{sec:bdydef}.  Section~\ref{sec:discussion} concludes with a discussion of our results and avenues for further investigation.  An appendix lays out our conventions on $\sltwo$ and $\sutwo$ current algebra following~\rcite{Balthazar:2021xeh}.



\section{Classical open strings on \texorpdfstring{$AdS_2$}{} branes}
\label{sec:classical}

While our ultimate interest is in the stringy $k<1$ regime of $AdS_3$, it is useful to consider the semiclassical limit of large $k$ to get a heuristic understanding of string dynamics in the presence of an $AdS_2$ D-brane in $AdS_3$.
Our discussion follows that of~\rcite{Martinec:2019wzw}.

\begin{figure}[ht]
\centering
\includegraphics[width=.4\textwidth]{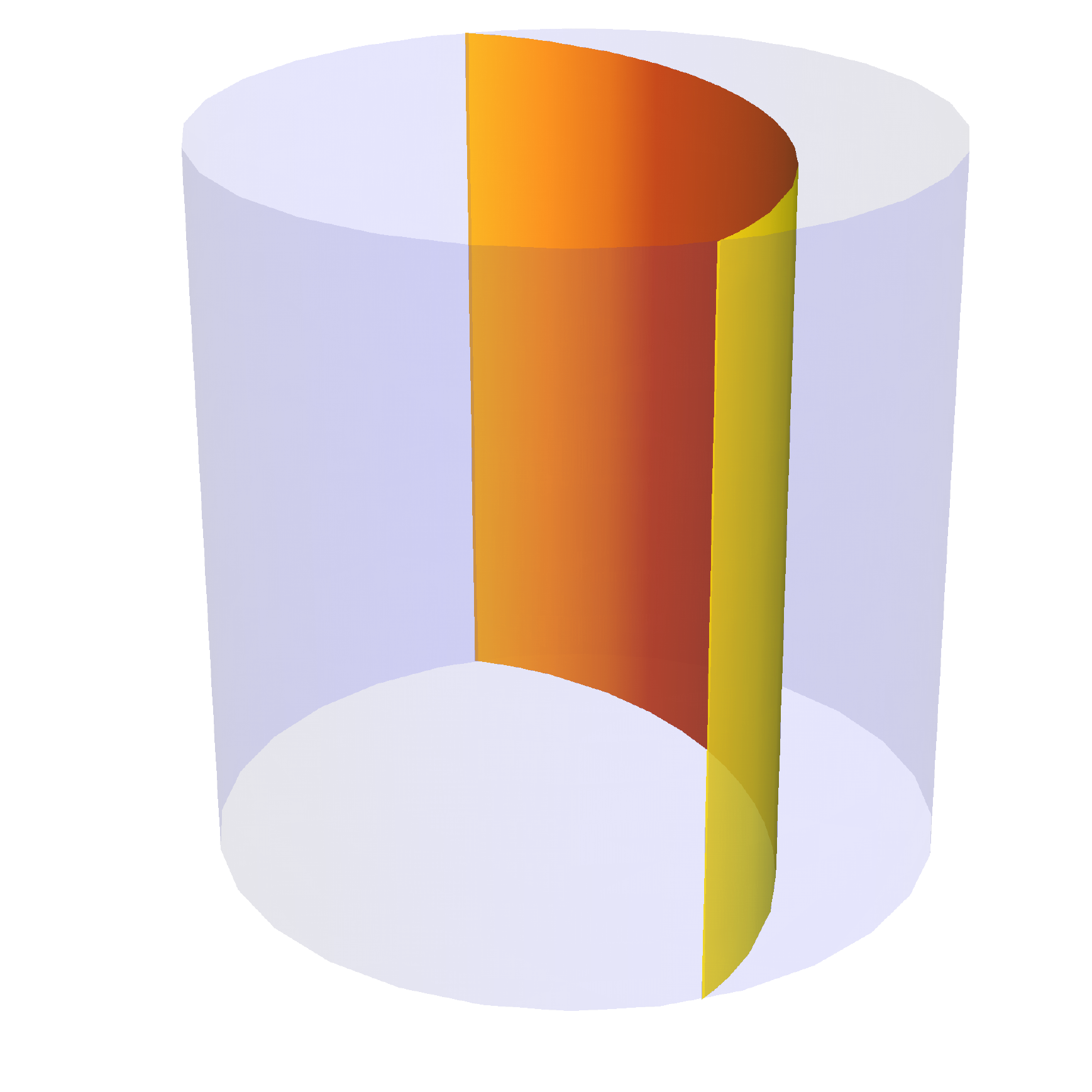}
\caption{\it An $AdS_2$ brane in $AdS_3$.}
\label{fig:AdS2brane}
\end{figure}

Classical solutions to the WZW model take the form
\eqn[classoln]{
g(\xi) = g_\ell(\xi_-) g_r( \xi_+) ~,
}
with $\xi_\pm=\xi_0\pm\xi_1$, and the left and right currents given by
\eqn[currents]{
j^a\equiv j_-^a(\xi_-) = \tr[(-it^a)g^{-1} \partial_+ g ]
~~,~~~~
\bar j^a\equiv j_+^a(\xi_+) = \tr[(-it^a)(\partial_+ g) g^{-1}] ~.
}
For open strings, the symmetry-preserving boundary conditions preserve a diagonal subgroup $G_{\rm diag}$ of the global bulk symmetry $G_L\times G_R$ on the boundary.  Symmetry-preserving branes thus lie along a (twisted) conjugacy class of the group 
\eqn[conjclass]{
\cC_G(f,\Omega) \equiv \big\{ g f \Omega(g^{-1})~,~~ g\in G\big\}
}
where $f\in G$ is a fixed group element and $\Omega$ is an automorphism of $G$.
The currents obey the boundary condition
\eqn[currentBC]{
\bar j=-\Omega( j)
}
Concretely, for $AdS_2$ branes within $\sltwo$ (see figure~\ref{fig:AdS2brane}), one has 
\eqn[SL2 outer auto]{
\Omega(g) = \sigma_1 g\sigma_1 
}
(which is an outer automorphism),
and the $AdS_2$ conjugacy class is specified by
\eqn[SL2 conj class]{
\tr[\sigma_1 g] = 2\sinh\mu ~.
}
The Euler angle parametrization of $SU(1,1)\simeq \sltwo$ 
\eqn[Euler]{
g = e^{\frac i2(\tau-\sigma)\sigma_3}\,e^{\rho \sigma_1}\, e^{\frac i2(\tau+\sigma)\sigma_3}
= \left(\begin{matrix} e^{i\tau}\cosh\rho ~&~ e^{-i\sigma}\sinh\rho \\
e^{-i\sigma}\sinh\rho  ~&~ e^{-i\tau}\cosh\rho  \end{matrix} \right)
}
provides standard global coordinates for $AdS_3$,
in which the metric is 
\eqn[adsmetric]{
ds^2=-\frac k2\tr[dg\,dg^{-1}]= k\big(d\rho^2-\cosh^2\!\rho\, d\tau^2+\sinh^2\!\rho\, d\sigma^2\big) ~.
}
Then the $AdS_2$ brane worldvolume is the codimension one surface
\eqn[conjclasscoords]{
\tr[\sigma_1 g] = 2\cos\sigma\,\sinh\rho = 2\sinh\mu ~.
}
Note that the $AdS_2$ brane hits the $AdS_3$ boundary $\rho\to\infty$ at $\sigma=\pm\frac\pi2$.  For nonzero $\mu$ the brane bends so as to avoid passing through the center of $AdS_3$, reaching a minimum radius $\rho=\mu$ at $\sigma=0$; for $\mu=0$ the brane runs straight across, passing through the center of $AdS_3$ at $\rho=0$.

%
\vspace{1mm}
%

For $\sutwo$, all automorphisms are inner; the conjugacy class describing a symmetry-preserving D-brane is an $\bS^2\subset\bS^3$.  In the Euler angle coordinates analogous to~\eqref{Euler}, the $\bS^3$ metric is
\eqn[sutwometric]{
ds^2 = n\big(d\theta^2 + \sin^2\theta\, d\phi^2 + \cos^2\theta\, d\psi^2\big) ~,
}
and the $\bS^2$ brane worldvolume is
\eqn[S2brane]{
\tr[g_\su] = 2\cos\theta\,\cos\psi = 2 \cos\mu' ~.
}
The solutions to this equation parametrize the polar direction of the $\bS^2$, while the $\phi$ circle provides the azimuthal direction, which degenerates at the two poles $\theta=0,\psi=\pm\mu'$.
Changing the automorphism $\Omega$ changes the orientation of $\bS^2\subset\bS^3$. 

The $\bS^2$ worldvolume carries $n_{D1}$ units of magnetic flux;%
\footnote{As a consequence, the size of the $\bS^2$ is quantized as $\mu'=2\pi n_{D1}/n$.}
altogether, the $AdS_2\times\bS^2$ worldvolume describes a dipolar D3-brane which is a bound state of $n_{D1}$ D1-branes puffed out into a sphere by the Myers effect~\rcite{Myers:1999ps}.  For $n_{D1}=1$ one has the trivial conjugacy class in $\sutwo$, and the brane is pointlike.  In the application to $k<1$ string theory, the $\bS^3$ is ``squashed'' by a marginal $\int\! J^3_\su\bar J^3_\su$ deformation as described in~\rcite{Balthazar:2021xeh}, which changes the metric to~\rcite{Hassan:1992gi,Giveon:1993ph}
\eqn[squashsphere]{
ds^2 = n\Big( d\theta^2 + \frac{R^2\cos^2\theta\, d\psi^2 + \sin^2\theta\,d\phi^2}{\cos^2\theta+R^2\sin^2\theta} \Big)
}
This deformation also squashes the $\bS^2$ worldvolume of the D-brane~\rcite{Forste:2001gn,Fredenhagen:2006dn}.%
\footnote{Because the deformation breaks $\sutwo$ down to $\uone$, one no longer has a moduli space of orientations of the brane in the ambient space~\rcite{Fredenhagen:2006dn}.}
One way to think about this deformation is to factorize the $\sutwo$ WZW model as $\big(\frac\sutwo\uone\times\uone\big)/\bZ_{n}$; the squashing deformation then acts to change the radius of the $\uone$ factor.
Equation~\eqref{S2brane} describes a line segment in the coset geometry which is the polar direction on the $\bS^2$ with the squashed $\uone$ parametrized by $\psi$ being the azimuthal direction.  The effect of this squashing on the operator spectrum of the quantum theory is described in the Appendix.

There are other symmetry-preserving D-branes preserving a diagonal subgroup of $\sltwo_L\times\sltwo_R$, as well as ``symmetry-breaking'' branes that preserve only a $\uone$ subgroup; for an overview and applications to $AdS_3\times\bS^3$ string theory, see for instance~\rcite{Martinec:2019wzw}.  We will briefly return to these additional possibilities in the discussion section.


\subsection{Classical closed strings}

The simplest classical solutions are point-like closed strings travelling geodesics in $AdS_3$,
\eqn[gengeodesic0]{
\g_{\alpha,\rho_-,\rho_+} = 
\Big( e^{\half\rho_-\sigma_1}\,e^{\frac i2\alpha \xi_-\sigma_3}\Big) 
\Big( e^{\frac i2\alpha\xi_+\sigma_3}\,e^{-\half\rho_+\sigma_1}\Big)  ~.
}
Recalling the Euler angle parametrization~\eqref{Euler},
one sees that this classical solution describes unexcited (pointlike) strings whose center of mass travels an elliptical trajectory oscillating between inner radius $\rho_\smile$ and outer radius $\rho^\frown$ (where $\rho^\pm=\rho^\frown\pm\rho_\smile$)
\eqn[ellipse]{
\sinh^2\rho = \cos^2(\alpha\xi_0) \,\sinh^2\rho_\smile + \sin^2(\alpha\xi_0)\sinh^2\rho^\frown ~.
}
In particular, for $\rho_\pm=0$ one has a string sitting in the center of $AdS_3$ travelling forward in time (for $\alpha>0$).

Classical spectral flow is the operation
\eqn[speclflowcl]{
\g^{(w)}_{\alpha,\rho_+,\rho_-} = e^{\frac i2 w \xi_-  \sigma_3}\, \g^{(w=0)}_{\alpha,\rho_+,\rho_-} \, e^{\frac i2 w \xi_+  \sigma_3}
}
spins the pointlike solution around in a circle, and so describes a circular string that gyrates around the origin between the same two limits $\rho_\smile$ and $\rho^\frown$.

\begin{figure}[ht]
\centering
  \begin{subfigure}[b]{0.35\textwidth}
    \includegraphics[width=\textwidth]{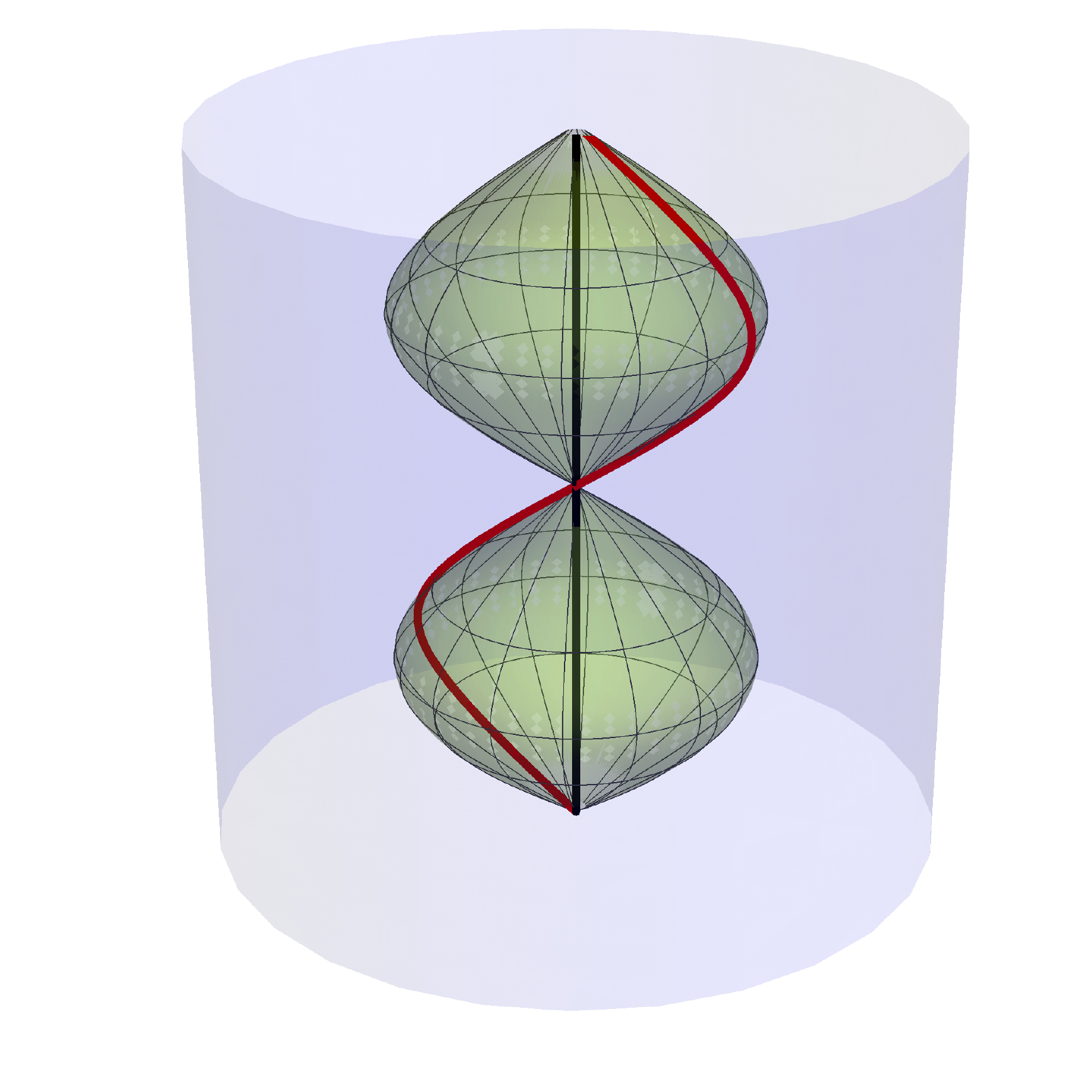}
    \caption{ }
    \label{fig:ShortClosed}
  \end{subfigure}
\qquad\qquad
  \begin{subfigure}[b]{0.35\textwidth}
      \hskip .5cm
    \includegraphics[width=\textwidth]{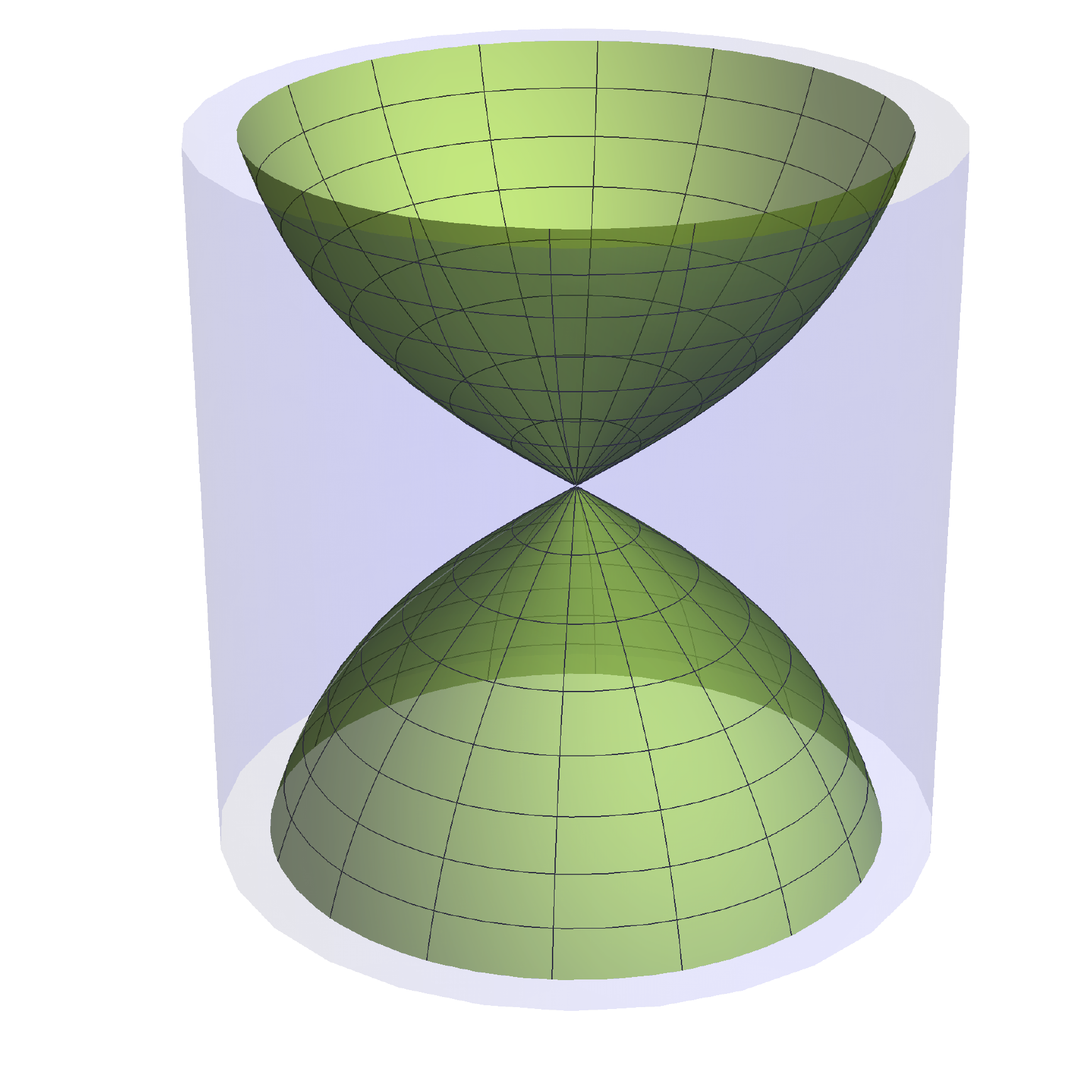}
    \caption{ }
    \label{fig:LongClosed}
  \end{subfigure}
\caption{\it 
Classical closed string solutions in $AdS_3$: (a) Highest weight states describe unexcited strings sitting at the origin (black trajectory), with zero mode descendants (red trajectory) describing strings oscillating about the origin having no oscillator excitations. Spectral flow spins a geodesic around $\rho=0$, describing strings that ``wind'' azimuthally, with zero mode descendants exciting a breathing mode oscillation (green worldsheet).
(b) Spectral flow of spacelike geodesics describes scattering states where the wound string approaches the $AdS$ boundary in the far past/future.
}
\label{fig:ClosedStrings}
\end{figure}

FZZ duality in this semiclassical context relates this gyrating round string to a coherent oscillator excitation (of the first Fourier mode) on top of the geodesic motion~\eqref{gengeodesic0}.  The oscillating classical string solution of interest is
\begin{align}
\gosc^{(w=0)}_{\alpha,\rho_+,\rho_-}(\xi) &= \gosc_\ell(\xi_-) \gosc_r(\xi_+)
\nn\\[.3cm]
\gosc_\ell (\xi_-) &=  \exp\Bigl[\frac{\rho_-}2\bigl( e^{-i\xi_-} \sigma_+ + e^{i\xi_-} \sigma_-\bigr)\Bigr]
\cdot\exp\Bigl[\frac i2\alpha\,\xi_- \, \sigma_3 \Bigr]
\\[.3cm]
\gosc_r (\xi_+) &=  
\exp\Bigl[\frac i2\alpha\,\xi_+ \, \sigma_3 \Bigr]
\cdot\exp\Bigl[-\frac{\rho_+}2\bigl( e^{i\xi_+} \sigma_+ + e^{-i\xi_+} \sigma_-\bigr)\Bigr]
\nn
\end{align}
One finds that if one takes the conjugate state of $g_{\rm osc}$ (the classical equivalent of exchanging $\cD^+\leftrightarrow\cD^-$), one arrives at a trajectory in the loop group of $\sltwo$ that is precisely the same as $\g^{w=-1}_{1-\alpha,\rho_+,\rho_-}$ :
\be
\label{semiclassicalFZZ}
\big[\gosc^{(w=0)}_{\alpha,\rho_+,\rho_-}(\xi)\big]^\dagger
= \g^{(w=-1)}_{1-\alpha,\rho_+,\rho_-}(\xi) ~~.
\ee
This equivalence is precisely the realization of FZZ duality on semiclassical coherent states!
Note that we can also spectral flow this relation.  The limit $\rho_\pm\to0$ recovers the primary highest weight states which sit at $\rho=0$ and travel up the $\tau$ direction.  Thus we see that at the classical level, FZZ duality is simply a global identification of coordinates in the loop group.

For an overview of spectral flow and FZZ duality in the quantum theory, see the Appendix.


\subsection{Classical open strings}

The classical solutions for open strings can again be constructed as factorized products, with the restriction that the string endpoints at $\xi_1=\pm\frac\pi2$ lie on the D-brane worldvolume.  The analogue of the unexcited closed string sitting statically in the center of $AdS_3$ is an open string statically stretched from the D-brane back to itself.  In general this string is not pointlike because there is an electric field on the D-brane worldvolume which induces a dipole moment (stretching apart the oppositely charged string endpoints) when $\mu\ne0$.  The static solution is
\eqn[staticopen]{
\g_\supset = \Big(e^{\frac i2 \alpha \xi_- \sigma_3}\, e^{\half\beta\sigma_1} \Big)
\Big(e^{\half\beta\sigma_1}\,e^{\frac i2\alpha \xi_+ \sigma_3} \Big)
= \left(
\begin{matrix} e^{i\alpha\xi_0}\cosh\beta ~&~ e^{-i\alpha\xi_1}\sinh\beta \\
e^{i\alpha\xi_1}\sinh\beta ~&~ e^{-i\alpha\xi_0}\cosh\beta \end{matrix}
\right)
}
The endpoints of the string lie on the $AdS_2$ brane provided
\eqn[openended]{
\cos(\alpha\pi/2)\,\sinh\beta = \sinh\mu ~.
}
When $\alpha=0$ the string is pointlike and lies at the center of the $AdS_2$ brane; as $\alpha$ increases the string stretches as the radial location of the endpoints moves out to larger radius, until at $\alpha=1$ the string endpoints lie at the boundary of $AdS_2\subset AdS_3$.  Due to the ambient $B$-field, the string does not cut straight across $AdS_3$ between the endpoints; rather, the string winds at fixed radius $\rho=\beta$ (at larger radius than the $AdS_2$ brane, on the convex side of the brane).

\begin{figure}[ht]
\centering
  \begin{subfigure}[b]{0.3\textwidth}
    \includegraphics[width=\textwidth]{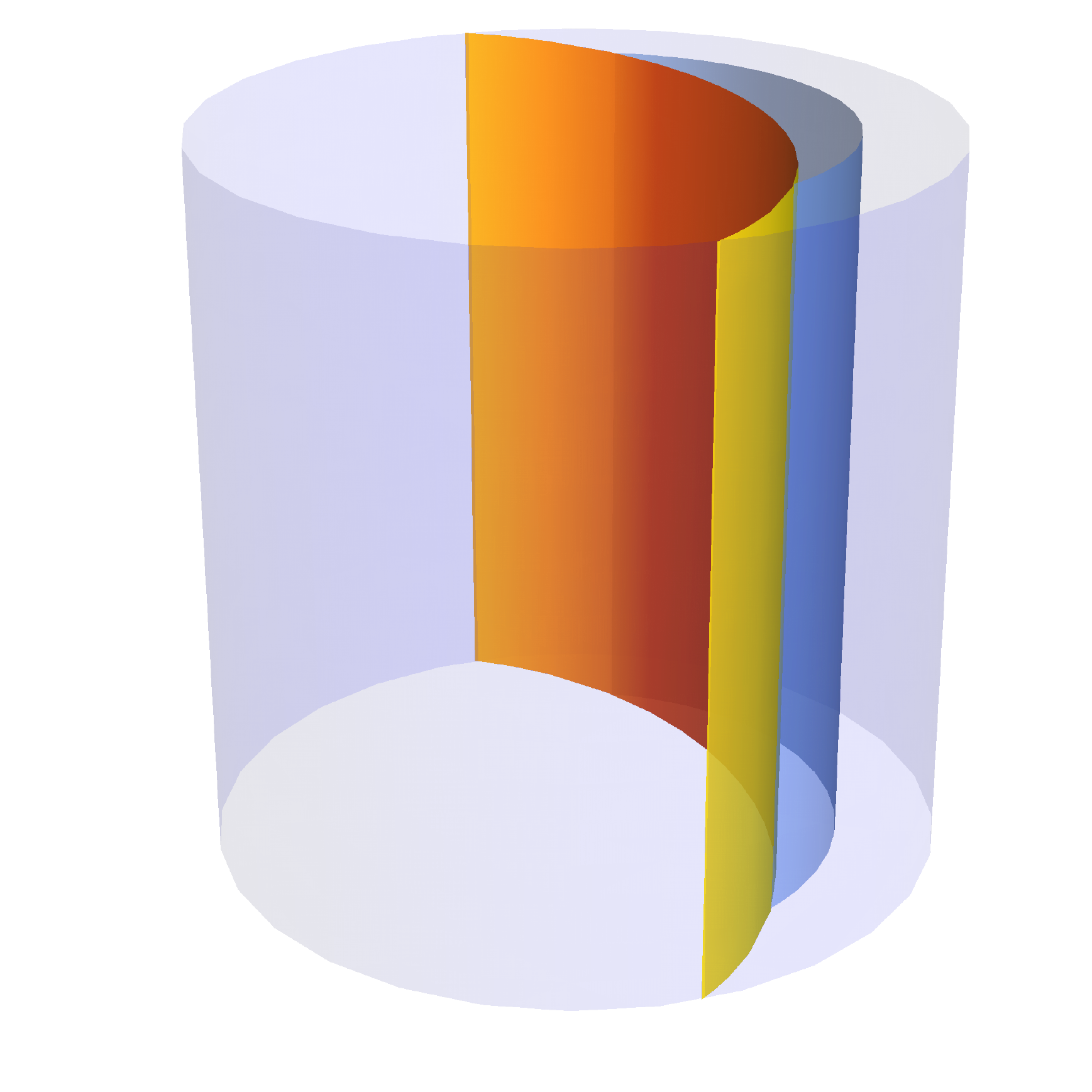}
    \caption{ }
    \label{fig:OpenString-out}
  \end{subfigure}
\qquad\qquad
  \begin{subfigure}[b]{0.3\textwidth}
      \hskip .5cm
    \includegraphics[width=\textwidth]{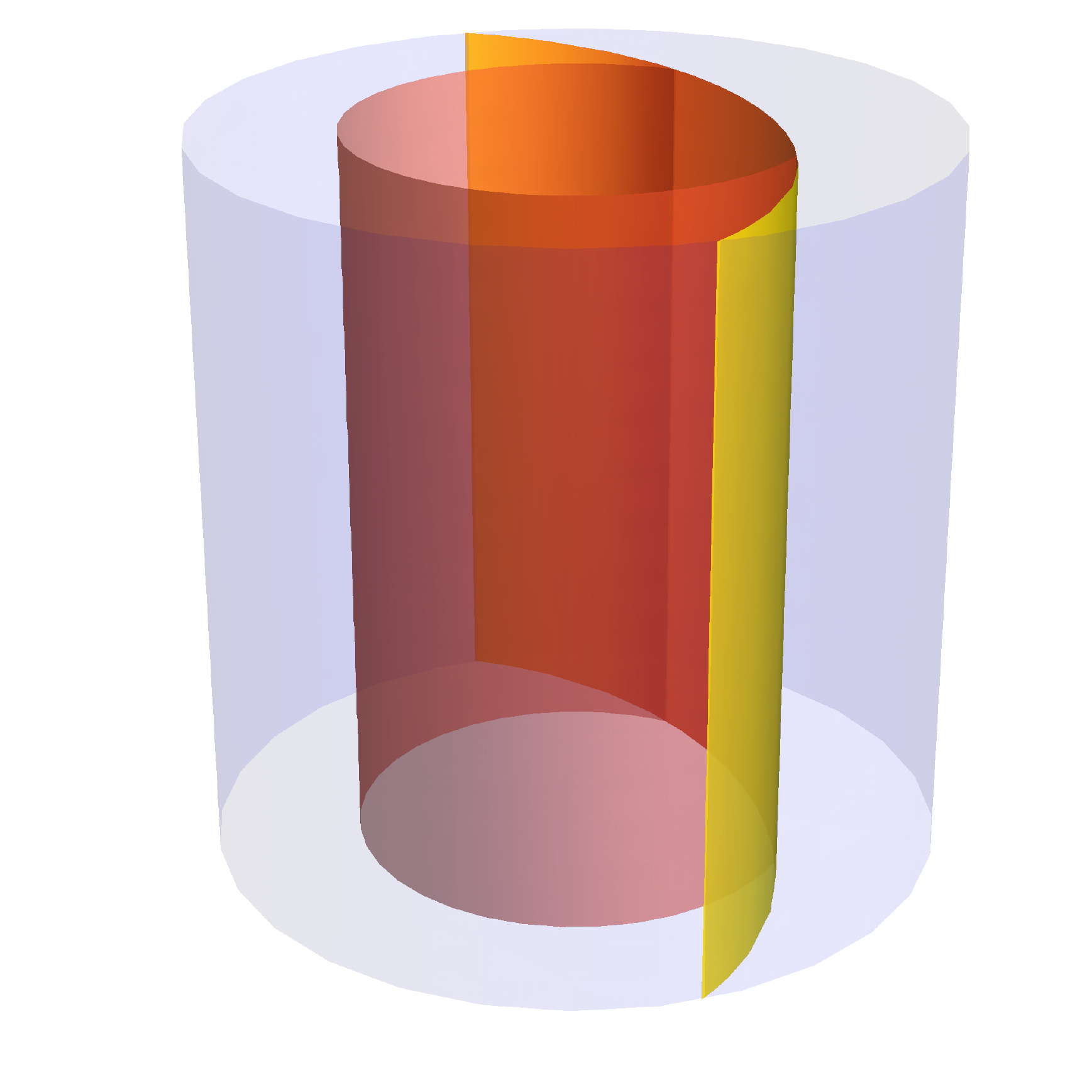}
    \caption{ }
    \label{fig:OpenString-in}
  \end{subfigure}
\caption{\it 
Classical string solutions for static open strings attached to an $AdS_2$ brane.  
The string can stretch on either (a) the convex side, or (b) the concave side, of the brane.
}
\label{fig:OpenStrings}
\end{figure}

Similarly, there is a second static solution which stretches statically in the opposite direction, toward the concave side of the brane
\eqna[otherstatic]{
\g_\subset &= \Big(e^{\frac i2(\pi+(1+\alpha) \xi_- )\sigma_3}\, e^{\half\beta\sigma_1} \Big)
\Big(e^{\half\beta\sigma_1}\,e^{\frac i2(-\pi+(1+\alpha) \xi_+ ) \sigma_3} \Big)
\\[.2cm]
&= \left(
\begin{matrix} ~e^{i(1+\alpha)\xi_0}\cosh\beta ~&~ -e^{-i(1+\alpha)\xi_1}\sinh\beta \\
-e^{i(1+\alpha)\xi_1}\sinh\beta ~&~ ~e^{-i(1+\alpha)\xi_0}\cosh\beta \end{matrix}
\right) ~,
}
with
\eqn[openended2]{
\cos[(1+\alpha)\pi/2]\,\sinh\beta = \sinh\mu ~.
}
Note that the midpoint of the brane at $\xi_1=0$ lies at $\sigma=\pi$, so indeed it pokes out on the concave side of the $AdS_2$ brane.  As we see from figure~\ref{fig:OpenString-in}, in~\eqref{otherstatic} the open string ends are being stretched apart by the bulk winding instead -- the dipole orientation is opposite to the worldvolume electric field.  There is a minimum energy cost for such a string, due to the extra winding in the bulk and the minimum radius of that winding in order to have a classical solution.  This lowest energy solution has $\alpha=0$; both ends of the string lie at the midpoint $\rho=\mu,\sigma=0$, and the string makes a complete circuit of the azimuthal circle.  On the other hand, increasing $\alpha$ we again find the string expands out toward the $Ads_3$ boundary, which it reaches as $\alpha\to 1$; as it does so, the total amount of winding decreases until at $\alpha=1$ it again winds halfway around $AdS_3$.

Physically, in figure~\ref{fig:OpenString-out} the bulk of the string is trying to draw the endpoints closer together, but as $\alpha$ increases the Lorentz force due to the B-field
\eqn[Lorentzforce]{
F_\rho = H_{\rho\tau\sigma}\partial_{\xi_0}\tau \partial_{\xi_1}\sigma}
reduces the effective tension of the string, allowing the worldvolume electric field on the $AdS_2$ brane to push the endpoints farther apart.  On the other hand, in figure~\ref{fig:OpenString-in} the bulk of the string is trying to pull the endpoints farther apart; because the orientation of the winding is the same, the dipole made by the string endpoints is aligned {\it against} the worldvolume electric field and so again the force is balanced between the worldsheet bulk and boundary contributions.

Note that, in terms of target space energy $\alpha=p_\tau\sim (2j-1)/k$, the solution~\eqref{staticopen} covers the range $0\le\alpha\le1$, while~\eqref{otherstatic} covers the range $1\le\alpha\le2$.  Spectral flow by $w\in 2\bZ$ generates additional wound and bound strings that circle $w/2$ times around the bulk azimuthally before reattaching to the brane.

The worldsheet energy density 
\eqn[Hamiltonian]{
\cE = -\frac k4\Big(\tr\big[\partial_0 g \partial_0 g^{-1}\big]
+\tr\big[\partial_1 g\partial_1 g^{-1}\big] \Big)
}
has the value $\cE=-k\alpha^2/4$ for~\eqref{staticopen}, and $\cE=-k(1+\alpha)^2/4$ for~\eqref{otherstatic}.  Note that $0\le\alpha<1$ for these solutions, so the ranges of energies don't overlap, with the strings winding on the concave side of the brane being more deeply bound.

\begin{figure}[ht]
\centering
    \includegraphics[width=.4\textwidth]{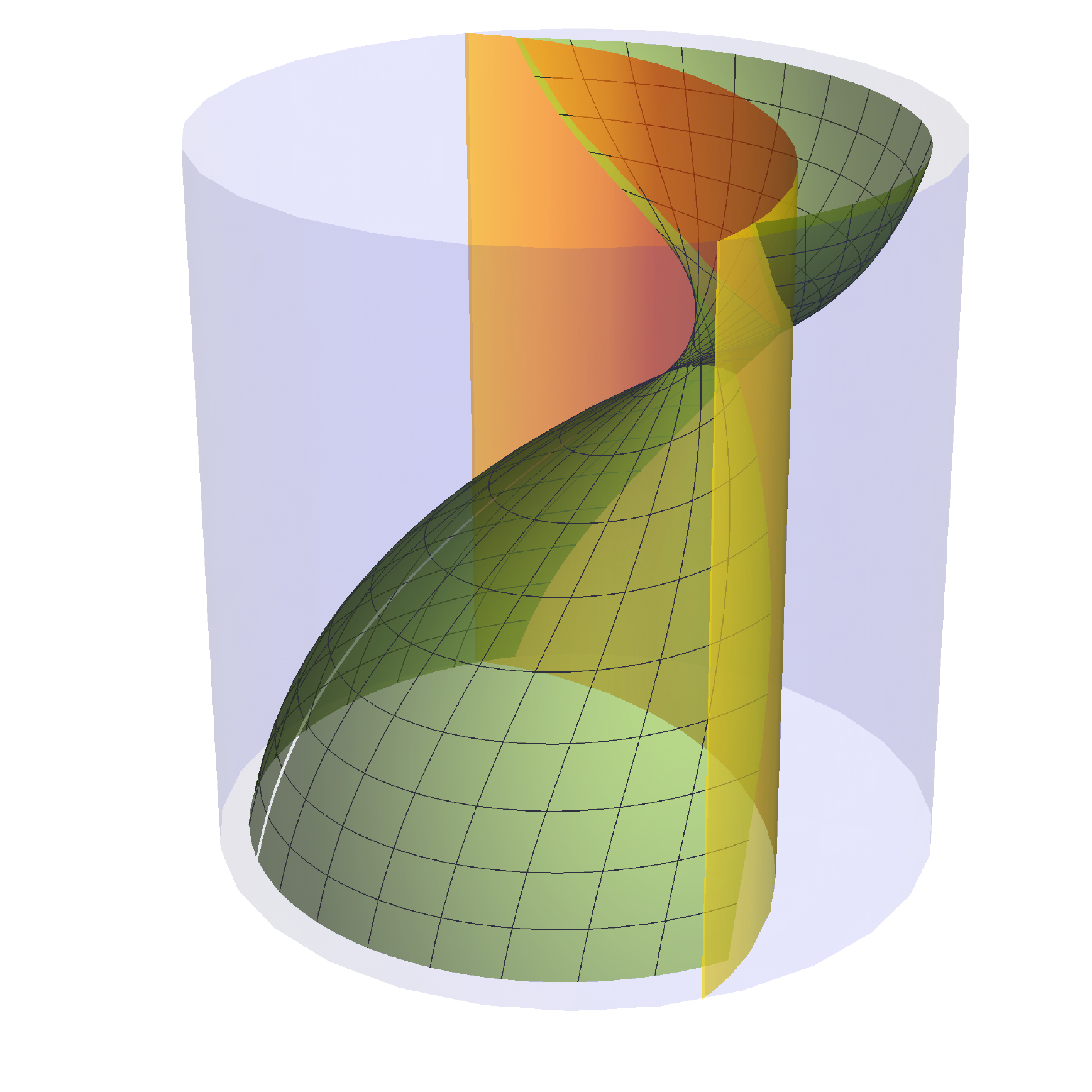}
\caption{\it 
The primitive classical long open strings initially wind halfway around one side of $AdS_3$ in the far past, and end up winding halfway around the opposite side in the far future.  
}
\label{fig:LongOpen}
\end{figure}

A class of long open string solutions
is given by
\eqna[longopen1]{
g &= \Big( e^{\frac i2(\frac\pi2+w\xi_-)\sigma_3} \, e^{\half \alpha \xi_- \sigma_2} \, e^{\frac i2 \beta\sigma_3} \Big)
\Big( e^{\frac i2 \beta\sigma_3} \, e^{\half \alpha \xi_+ \sigma_2} \, e^{\frac i2(-\frac\pi2+w\xi_+)\sigma_3} \Big)
\\[.2cm]
&= \left(\begin{matrix}
e^{iw\xi_0}\big(\cos\beta\,\ch(\alpha\xi_0)+i\sin\beta\,\ch(\alpha\xi_1)\big) 
~&~
e^{-iw\xi_1}\big(\cos\beta\,\sh(\alpha\xi_0) + i\sin\beta\,\sh(\alpha\xi_1)\big)
\\
e^{iw\xi_1}\big(\cos\beta\,\sh(\alpha\xi_0) - i\sin\beta\,\sh(\alpha\xi_1)\big)
~&~
e^{-iw\xi_0}\big(\cos\beta\,\ch(\alpha\xi_0)-i\sin\beta\,\ch(\alpha\xi_1)\big)
\end{matrix}\right)  ~.
}
One has 
\eqn[conjclass1]{
\tr[g\sigma_1] = 2\big(
\sin(w\xi_1)\,\cos\beta\,\sinh(\alpha\xi_0) 
- \cos(w\xi_1)\,\sin\beta\,\sinh(\alpha\xi_1) \big)
}
and so for $w\in 2\bZ+1$, the endpoints at $\xi_1=\pm\frac\pi2$ lie along the $AdS_2$ brane provided
\eqn[longstringends1]{
\sin\beta\,\sinh(\alpha\pi/2) = \sinh\mu 
}
(thus once one fixes an asymptotic radial momentum $\alpha$ and winding $w$, the solution is completely determined).
In these solutions, the minimum radius reached by the brane (at $\xi_0=\xi_1=0$ is $\rho=0$.
Note that a string that starts off on the convex side of the brane in the far past ends up on the concave side in the far future, and vice-versa.  An example of such long open strings is depicted in figure~\ref{fig:LongOpen}.  The worldsheet energy of this solution is%
\footnote{Quantum mechanically, we identify $\alpha=(2j-1)/k$, and there is a Casimir contribution to the energy $\delta\cE=-1/4k$, in order to match the current algebra spectrum $\cE=-j(j-1)/k+kw^2/4$.}
\eqn[longenergy]{
\cE = \frac k4\big( \alpha^2-w^2\big) ~.
}

\begin{figure}[ht]
\centering
    \includegraphics[width=.4\textwidth]{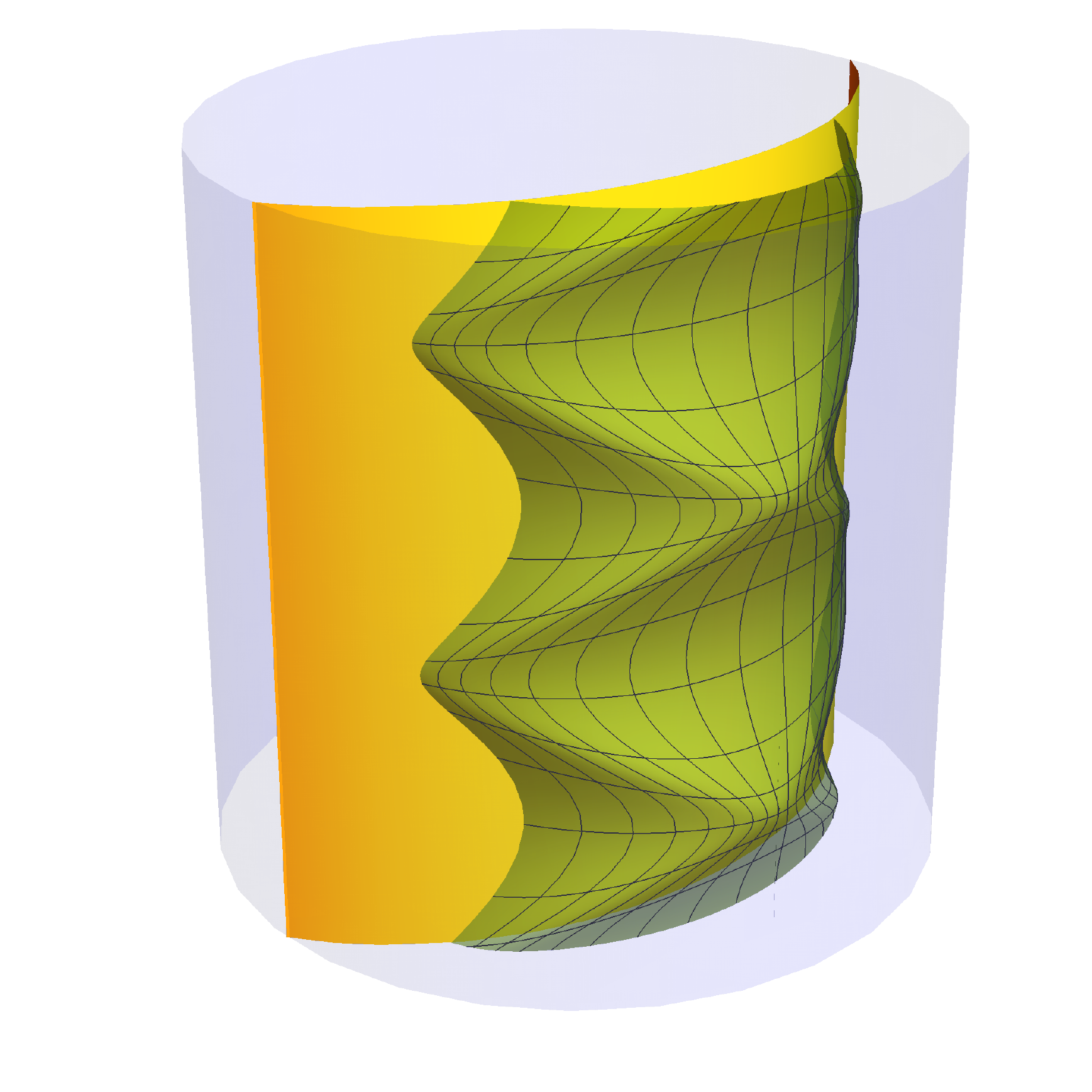}
\caption{\it 
Oscillator excitation above the ground state.  
}
\label{fig:wiggles}
\end{figure}

The above solutions are essentially ground states of the string, in the sense that oscillators are not excited.  One can excite oscillations in the following way: Consider a general chiral group-valued function $h(\xi)$, and act on the solution via
\eqn[excitedstring]{
g \longrightarrow h(\xi_+)\,g\,\Omega\big(h^{-1}(\xi_-)\big)  ~.
}
This is precisely the diagonal action that preserves the boundary condition, but in general the resulting classical solution has additional wiggles coming from those of $h(\xi)$.  For instance, the particular choice
\eqn[hwiggle]{h(\xi) = \exp\big[a\big(\cos \xi\,\sigma_1+ \sin \xi\, \sigma_2\big)\big] }
leads to the excited string depicted in figure~\ref{fig:wiggles}.

As for closed strings, classical FZZ duality for open strings is a global identification of elements of the loop group, in which exponentials of generators with only the zero mode coherently excited yield precisely the same element of the loop group built out of exponentials of generators with the first oscillator mode excited
\eqna[goscopen]{
\gosc_{\supset}(\alpha,\beta;\xi_+,\xi_-) &= \gosc_\ell(\alpha,\beta;\xi_-)\, \gosc_r(\alpha,\beta;\xi_+)
\\[.2cm]
\gosc_\ell(\alpha,\beta;\xi_-) &=
\Big( e^{\frac i2 \alpha\xi_-\sigma_3}\, \exp\Big[\hf\beta\big(e^{-i\xi_-} \sigma_+ + e^{+i\xi_-}\sigma_-\big)]\,e^{-\frac i2 \xi_-\sigma_3} \Big)
\\[.2cm]
\gosc_r(\alpha,\beta;\xi_+) &=
\Big(e^{-\frac i2 \xi_+\sigma_3} \, \exp\Big[\hf\beta\big(e^{+i\xi_-} \sigma_+ + e^{-i\xi_-}\sigma_-\big)]\, e^{\frac i2 \alpha\xi_+\sigma_3} \Big)~.
}
Evaluating the product of group elements, one has
\eqn[goscopenmatrix]{
\left(\begin{matrix}
e^{i(\alpha-1)\xi_0}\cosh\beta ~&~ e^{-i(\alpha-1)\xi_1}\sinh\beta \\
e^{i(\alpha-1)\xi_1}\sinh\beta ~&~ e^{-i(\alpha-1)\xi_0}\cosh\beta 
\end{matrix}\right) ~.
}
Comparing the above to the group element $\g$ in~\eqref{staticopen} where only the zero modes are excited in the exponents, we have 
\eqn[FZZopen]{
\g_\supset(\alpha,\beta) = [\gosc_\supset(1-\alpha,\beta)]^\dagger ~,
}
which is the classical statement of FZZ duality for open string ground states~-- the adjoint relates discrete series representations $\cD^+\leftrightarrow \cD^-$, and our conventions determine $\alpha= (2j-1)/k$ so that $j\to \tilde\jmath=k/2+1-j$.  

Similarly, the oscillator-excited FZZ dual for the other class of static solutions~\eqref{otherstatic} is given by
\eqna[othergoscopen]{
\gosc_{\subset}(\alpha,\beta;\xi_+,\xi_-) &= \gosc_\ell(\alpha,\beta;\xi_-)\, \gosc_r(\alpha,\beta;\xi_+)
\\[.2cm]
\gosc_\ell(\alpha,\beta;\xi_-) &=
\Big( e^{\frac i2(\pi+ \alpha\xi_-)\sigma_3}\, \exp\Big[\hf\beta\big(e^{-i\xi_-} \sigma_+ + e^{+i\xi_-}\sigma_-\big)]\,e^{-\frac i2 \xi_-\sigma_3} \Big)
\\[.2cm]
\gosc_r(\alpha,\beta;\xi_+) &=
\Big(e^{-\frac i2 \xi_+\sigma_3} \, \exp\Big[\hf\beta\big(e^{+i\xi_-} \sigma_+ + e^{-i\xi_-}\sigma_-\big)]\, e^{\frac i2(-\pi+ \alpha\xi_+)\sigma_3} \Big)~,
}
with the relation
\eqn[otherFZZopen]{
\g_\subset(\alpha,\beta) = [\gosc_\subset(-\alpha,\beta)]^\dagger ~,
}

This classical FZZ duality for open strings is a bit different in some respects than its closed string counterpart described in~\rcite{Martinec:2020gkv}.  In the closed string sector, the ground state string is pointlike semiclassically.  Well-defined trajectories such as the red and black trajectories of figure~\ref{fig:ShortClosed} are coherent state wavepackets well localized on the AdS scale.  
Quantum mechanically, the wavefunction
\eqn[ptclewavefn]{
\psi(\rho)\sim (\cosh\rho)^{-2j}}
is indeed becoming more and more concentrated for large $j$, and the localized particle-like behavior arises for $j\gg1$. On the other hand, the corresponding FZZ dual winding condensate wavefunction has $\tilde\jmath=\half k+1-j$ and so is well localized already for small $j$, but becomes more and more spread out for large $j$, until at $j\sim k/2$ the winding wavefunction is spread out over the $AdS$ scale and the particle wavefunction is concentrated on the string scale; in this regime we are near the threshold of the continuum of long strings.  For open strings, one instead has the phenomenon that even classically, the string spreads out as $\alpha=(2j-1)/k$ increases, due to the interaction with the worldvolume electric field, as we see in figure~\ref{fig:OpenStrings}.  Only for vanishing electric field $\mu=0$ does the situation resemble that of the closed strings ($\mu=0$ forces $\rho=\beta=0$ in equations~\eqref{openended}, \eqref{openended2} for the radial location of the unexcited string), with all the ground state strings being pointlike in the semiclassical approximation.


\section{The quantum theory}
\label{sec:quantumcomments}

The semiclassical analysis above provides a conceptual basis for the exact quantum theory, whose solution systematically exploits the structure of $\sltwo$ current algebra and its representation theory%
~\rcite{Lee:2001xe,Lee:2001gh,Ponsot:2001gt,Giveon:2001uq,Israel:2005ek,Hosomichi:2006pz}.  After a brief overview of the closed string spectrum following~\rcite{Balthazar:2021xeh}, we describe briefly the boundary operators on an $AdS_2$ brane in the worlsheet theory, whose structure is completely analogous to that of bulk operators reviewed in the Appendix.  We then apply this structure to the description of open string vertex operators.


\subsection{Overview of the closed string spectrum}
\label{sec:closedspec}

An extensive analysis of the closed string spectrum of the worldsheet theory was given in~\rcite{Balthazar:2021xeh}, to which we refer for details.  We describe here a few highlights.  

Spacetime supersymmetry requires the squashing deformation mentioned above, with parameter $R=\sqrt{n+1}$.  For the present application, it is simplest to work with the type IIB theory since it is left-right symmetric on the worldsheet, and so worldsheet boundary conditions are more straightforward to describe.  
The squashing is best described by factoring
\eqn[cosetdecomp]{
\sutwo_n=\biggl(\frac{\sutwo_n}\uone\times\uone\biggr)/\bZ_n
}
into the $\cN=2$ supersymmetric coset theory $\frac\sutwo\uone$ (which we will also refer to by its various other characterizations as a Landau-Ginsburg model, or as a parafermion theory) and the remaining $\uone$ on which the squashing transformation acts. 

There are GSO projections for both $\cN=(2,0)$ type IIA and $\cN=(2,2)$ type IIB string theories.  The chiral GSO projection is 
\eqna[chiralGSO]{
(-1)^F\,\Omega &= 1 ~,
}
with the component contributions
\eqna[GSOphase]{
(-1)^F &= \exp\Big[i\pi\Big(-q_\varphi + \epsilon_\sl\eta_\sl + \epsilon_3\eta_3 + \epsilon_\su\eta_\su \Big)\Big]
\\[.2cm]
\Omega &=
\exp\Big[i\pi \Big( (m'\tight-\mbar') + (\eta_\su\tight-\bar\eta_\su) +\frac{n-2}2(w'\tight-\bar w') \Big) \Big] ~,
}
where $q_\varphi$ is the spinor ghost charge, $\eta_i$ are the bosonized fermion numbers, and $\epsilonb=(\epsilon_\sl,\epsilon_3,\epsilon_\su)=(-1,1,1)$.  This projection allows the spacetime supersymmetry and $R$-symmetry charges, given by contour integrals of
\eqna[susyops]{
S_r^\pm &= \exp\bigg[-\frac\varphi 2 +ir\big(H_\sl\mp H_3\big) \pm\frac {i\,a}2\, Z \pm \frac{i}{\sqrt{2k}}\,Y \bigg]~,
\\[.2cm]
J_R &= i\sqrt{2k}\, \partial Y~,
}
where $a=\sqrt{1-\frac2n}$, $r=\pm\half$, $Y$ bosonizes the total $\uone$ in the supersymmetric coset decomposition~\eqref{cosetdecomp}, $Z$ bosonizes the R-symmetry of the parafermion theory, and $H_\sl,H_3$ bosonize $\psi^\pm_\sl,\psi^3_\su\pm\psi^3_\sl$ (see the Appendix for specifics).

Vertex operators are built around a center-of-mass component 
\eqn[NSop]{
\Phi^{(w)}_{j;m,\mbar}\,\Lambda^{(w',\bar w')}_{j';m',\mbar'}\,e^{ip_Y Y+i\bar p_Y\bar Y}
}
describing a wavefunction in $AdS_3\times\sqsphere$, where $\Phi$ is an $\sltwo$ primary as described in Appendix~\ref{sec:sltwo}, $\Lambda$ is an $\frac\sutwo\uone$ coset primary as described in Appendix~\ref{sec:sutwo}, and the exponential is a $\uone$ primary whose momenta $p_Y,\bar p_Y$ are determined in terms of the quantum numbers of the parafermion by~\eqref{squonespec}, \eqref{paramrel}.  One then decorates this (for NS sector operators) with a polynomial in worldsheet currents and their superpartners for $\sltwo$ and $\uone$, and parafermions for $\frac\sutwo\uone$, as well as a ghost exponential $e^{-\varphi}$ if we are working in the $-1$ picture of the spinor ghosts (similarly for right-movers).  One then has to analyze the BRST constraints, and take linear combinations to find representatives of the BRST cohomology.  

For the Ramond sector, one needs a spin field $S$ of the $\sltwo\times\sutwo$ theory, processed through the squashing deformation and again decorated with the action of worldsheet currents, parafermions, \etc., as well as a ghost exponential $e^{-\varphi/2}$ for the $-1/2$ picture of the spinor ghosts (similarly for right-movers); and again, one needs to take linear combinations to find representatives of the BRST cohomology.

For real $j$, most of the usual BRST invariant string vertex operators describing perturbations of $AdS_3$ lie outside of the unitary range $\half<j<\half(k+1)$, and so do not correspond to normalizable string states; rather, they are Fourier modes of non-normalizable operators which are the worldsheet representatives of operators in the spacetime CFT (again see Appendix~\ref{sec:sltwo} for a sketch of the $\sltwo$ representation theory, and~\rcite{Balthazar:2021xeh} for a detailed construction of the low-lying spectrum).  Among the operators that do correspond to states are modes of the closed string ``tachyon'' that survive the noncritical type IIB GSO projection, \ie~\eqref{NSop} decorated by $e^{-\varphi-\bar\varphi}$, with $w=w'=\bar w'=0$ and
\eqn[pYtach]{
\left(p_Y,\bar p_Y\right)=\frac1{\sqrt{2n}}
\left[\frac{m'+\mbar'}{R} \pm \big(m'-\mbar' \big)R\right] ~,
}
and the squashing parameter set to $R=\sqrt{n+1}$ in the supersymmetric theory.  The GSO projection is satisfied for $m'-\mbar'\in2\bZ+1$, and one can check that despite the name, none of the modes surviving the projection are tachyonic for this value of $R$ (as required by supersymmetry).
The mass shell condition
\eqn[tachshell]{
-\frac{j(j-1)}{k} + \frac{j'(j'+1)-m'^2}{n} 
+ \frac{1}{4n}\Big[\frac{m'+\mbar'}{R}+(m'-\mbar')R\Big]^2 =\half ~.
}
determines the value of $j$.  One finds that $j'\tight=\half,m'\tight=-\mbar'\tight=\pm\half$ survives the GSO projection and has $j=\half+\frac1{\sqrt{n+1}}$ in the unitary range (all the other modes have $j>\half(k+1)$ and are thus non-normalizable).

The winding one sector $w=-1$ corresponds to the untwisted sector or block theory $\cM$ of the symmetric product in spacetime.  Consider the vertex operators
\eqn[blocktach]{
e^{-\varphi-\bar\varphi}\, e^{iH_\sl+i\bar H_\sl} \,
\Phi^{(-1)}_{j;m,\mbar}\, e^{i(p_Y Y+\bar p_Y\bar Y)} \, 
\Lambda_{j';m',\mbar'}^{(0,0)} \, 
~, 
}
where $H_\sl,\bar H_\sl$ bosonize the $\sltwo$ fermions $\psi^\pm_\sl$, $m=\bar{m}$ is found from the on-shell condition,
\eqn[mblocktach]{m=\frac{j(j-1)}{k}+\frac{k+2}{4}-\frac{j'(j'+1)-m'^2}{n}-\frac{p_Y^2}{2}~,}
and the momenta $(p_Y,\bar p_Y)$ are given in \eqref{pYtach}. The extra $\sltwo$ fermions $\psi^+_\sl\bar\psi^+_\sl=e^{iH_\sl+\bar iH_\sl}$ on left and right arise because the spectral flow is in the total $J^3_\sl$ and $\bar J^3_\sl$, and so flows both the bosonic winding and the fermion number.  This operator has spacetime dimension 
\eqn[hSTblocktach]{h_{\ST}=-m-1+\frac{k+2}{2}=-\frac{j(j-1)}{k}+\frac{k-2}{4}+\frac{j'(j'+1)-m'^2}{n}+\frac{p_Y^2}{2}~,}
with $\bar h_{\ST}=h_{\ST}$.
The corresponding operator in the block $\cM$ of the spacetime CFT is the general wave operator on $\bR^\phi\times \sqsphere$
\eqn[blocktachST]{e^{\beta\phi}\, e^{i\left(p_Y Y+\bar p_Y\bar Y\right)} \, 
\Lambda_{j';m',\mbar'}^{(0,0)}~,}
with $\beta$ related to $j$ by
\eqn[expII]{
\beta=-\frac{Q_\ell}{2} + \sqrt{\frac 2k}\Big(j-\half\Big) ~.
}
For further examples, including the construction of the chiral algebra of the block theory in the worldsheet representation, see~\rcite{Balthazar:2021xeh}.


\subsection{Boundary operators in \texorpdfstring{$SL(2,\bR)$}{}}
\label{sec:sltwobdyops}

The representation theory for boundary operators parallels the classical solutions above, as well as that of closed strings reviewed in Appendix~\ref{sec:sltwo}.  The diagonal $\sltwo$ symmetry preserved by the boundary conditions organizes the representations: Bound states lie in discrete series representations $\cD^\pm_{j;w}$ for $\half<j<\half(k+1)$, while scattering states lie in continuous series representations $\cC_{j,\alpha;w}$ for $j=\half+is$, $s\in\bR$, and $\alpha\in(0,1)$.  The spectral flow quantum number $w\in\bZ$ characterizes the various winding sectors and the action of the boundary currents on the associated vertex operators $\Psi^{(w)}_{j;m}$
\eqna[bdysl2reps]{
&j^3_\sl(z) \Psi^{(w)}_{j;m}(0)\sim \frac{m+\frac{k+2}{2}w}{z}\,\Psi^{(w)}_{j;m}(0)~,\cr
&j^\pm_\sl(z) \Psi^{(w)}_{j;m}(0)\sim \frac{m\mp (j-1)}{z^{\pm w+1}}\,\Psi^{(w)}_{j;m\pm1}(0) ~.
}
Note, however, that $w\in\bZ$ corresponds to an open string that winds $w/2$ times azimuthally.
The standard Sugawara construction of the stress tensor $T(z)=\frac1k{j^a j_a}$  determines the conformal dimension
\eqn[bdyopdim]{
h\big[\Psi^{(w)}_{j;m}\big] = -\frac{j(j-1)}k - mw - \frac{k+2}4 w^2 ~.
}

It is also useful to describe operators in a dual ``position-basis'', which for operators associated to states can be thought of as a coherent state representation (say in the winding zero sector)
\eqn[cohstate]{
\Psi_j^{(0)}(u;z) = e^{-uJ^-_0}\,\Psi^{(0)}_j(0;z)\,e^{uJ^-_0}
}
related to the ``momentum-basis'' of eigenstates of $J^3$ via
\eqna[Psiexpn]{
\Psi_{j;m}^{(0)} &= \int\! du\, u^{j+m-1} \, \Psi^{(0)}_{j}(u) 
}
where we have suppressed the $z$-dependence.
Semiclassically, the bulk $x$-basis operator $\Phi_j^{(0)}(x,\bar x)$ of~\eqref{scalingfns} has the form of a bulk-to-boundary propagator~-- a solution of the linearized bulk wave equation (an essential characteristic of a string vertex operator) with the property that it localizes asymptotically to a point on the boundary, see equation~\eqref{Phiasymp}.  The boundary operators $\Psi_j^{(0)}(0)$ are semiclassically the $AdS_2$ analogues~-- solutions of the $AdS_2$ wave equation that asymptotically localize on the $AdS_2$ boundary of the D-brane.

The coherent state parameters $(x,\bar x)$ of the $x$-basis parametrize the $AdS_3$ conformal boundary, and thus specify locations of operator insertions in the worldsheet representation of spacetime CFT correlators.  Similarly, $u$ parametrizes the $AdS_2$ conformal boundary, which is the location of a ``defect'' in the spacetime CFT dividing its coordinate domain into two halves.  For instance, in the Euclidean continuation of $AdS_3$ to the hyperbolic space $\bH_3^+$ with boundary $\bS^2$, the $AdS_2$ brane continues to a hemisphere ending in a boundary circle $\bS^1\subset\bS^2$.  Thus $x\in\bC$ parametrizes the boundary sphere, while $u\in\bR$ parametrizes a circle dividing that sphere into two halves which we take to be the upper and lower half-planes.  
In Lorentz signature, $x,\bar x$ analytically continue to null coordinates $x^\pm$ on the cylindrical conformal boundary of $AdS_3$, and $u$ is a timelike coordinate on the conformal boundary of $AdS_2\subset AdS_3$ (which has two components), which splits the boundary cylinder into two complementary half-cylinders.

In general, such a defect line in the spacetime CFT is not a boundary of the coordinate domain; rather, excitations impinging upon the defect can be partially transmitted and partially reflected~\rcite{Bachas:2001vj}, and correlators need not be reflection symmetric under $x\leftrightarrow \bar x$.  The worldsheet theory tells us that this more general situation applies here.  At leading order, strings mostly pass through the $AdS_2$ brane, interacting with it only at order $\gstr$.  Closed string operators $\Phi^{(0)}_j(x,\bar x)$ can be inserted on either the upper or lower half plane of $x$ (or on either half-cylinder of the Lorentzian version).  Furthermore, in the presence of a nonzero $\mu$, there is an asymmetry between the two sides of the $AdS_2$ brane and so correlators in general are not reflection symmetric (as one sees for instance in the closed string one-point-function on the disk, \cf~\rcite{Ponsot:2001gt}, equation 3.35).

\begin{figure}[ht]
\centering
    \includegraphics[width=.7\textwidth]{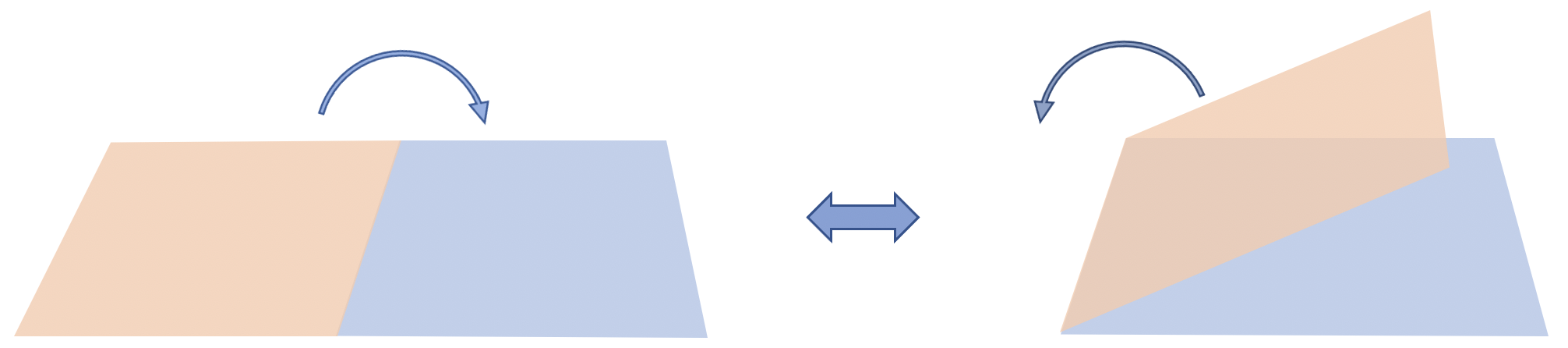}
\caption{\it 
Folding a defect CFT into a boundary CFT.  Reflection and transmission coefficients across the defect on the left figure, become a matrix of reflecting boundary conditions on the right figure.
}
\label{fig:unfold}
\end{figure}

One can always turn such a defect into a boundary condition in a boundary CFT by a folding trick which reflects the CFT $\cM$ in the LHP into a CFT $\widebar\cM$ in the UHP, see figure~\ref{fig:unfold}, with the reflection and transmission coefficients across the defect becoming a matrix of boundary conditions specifying how much of incident stress-energy in $\cM$ reflects back into $\cM$ versus transmits into $\widebar\cM$~\rcite{Bachas:2001vj}.  Indeed, we will adopt this perspective in constructing the spacetime CFT dual of $AdS_3$ string theory at $k<1$.


\subsection{Open string vertex operators}
\label{sec:bdyvertices}

It is important to remember that exponentials of free bosons have somewhat different properties on the boundary as compared to the bulk of the coordinate domain.  Due to the image term in the propagator of a free field $Y$
\eqn[FFT2pt]{
\big\langle Y(z_1,\bar z_1)\,Y(z_2,\bar z_2) \big\rangle = -\log|z_1-z_2|^2-\log|z_1-\bar z_2|^2 ~,
}
the stress tensor OPE with an exponential operator $e^{\alpha Y}$ is modified such that the conformal dimension of a boundary exponential is
\eqn[bdydim]{
h_\partial^{~} \big[e^{\alpha Y}\big] = -2\alpha^2-Q \alpha
}
as compared to that of a bulk exponential, for which the second term does not contribute to the bulk OPE and thus 
\eqn[bulkdim]{
h_{\rm bulk}[e^{\alpha Y}]=-\half \alpha(\alpha+Q) ~.
}
Thus, when considering spectral flow of a boundary operator, an exponential $e^{iwY/2}$ for $w\in \bZ$ implements an integer amount $w$ of spectral flow as far as $\sltwo$ representation theory is concerned, however the amount of winding that the string makes around the $AdS_3$ azimuthal direction is half as much, namely $w/2$ units of winding.  Note that the same halving of exponents will apply to the spacetime CFT in the presence of a boundary.

A highest weight boundary operator in the bosonic $\sltwo$ CFT is labelled~by
\eqn[sltwobdy]{
\Psi^{(w)}_{j} ~,
}
where the spectral flow quantum $w\in\bZ$ but the winding is $w/2$, and again the principal quantum number is restricted to $-\half<j<\half(k+1)$ for vertex operators associated to normalizable states in the discrete series $\cD^\pm_j$, while $j=\half+is$, $s\in\bR$ for vertex operators associated to delta-function normalizable states in the continuous series $\cC_{j;\alpha}$.%
\footnote{Continuous series representations carry an extra label $\alpha$ denoting the fractional part of $m-j$, which we will suppress.}

Highest weight open string vertex operators on $AdS_2$ branes lie in representations whose center of mass degree of freedom lies in a wavefunction on $AdS_2$.  Since $j^3=\bar j^3$ on the boundary, the zero mode representations in the Fourier basis are labeled by $(j,m)$ rather than $(j,m,\mbar)$.  In the $x$-basis, they depend on a single variable $u$ that parametrizes the location of the operator along the defect on the conformal boundary.

The analysis of the boundary vertex operator spectrum follows straighforwardly from the analysis of closed string vertex operators in~\rcite{Balthazar:2021xeh} sketched above.  Once again they have the general structure of a center of mass operator
\eqn[opencom]{
\Phi^{(w)}_{j;m}\,\Lambda^{(w')}_{j';m'}\,e^{ip_Y Y} ~,
}
with ghost factors (and a spin field for the Ramond sector),
decorated with oscillator excitations and subject to the BRST constraints and GSO projection.  We take the D-brane in $\sqsphere$ to be a squashed $\bS^2$ conjugacy class, labelled by an integer ${\jb'=0,\hf,1,...,\frac k2}$, for which the spectrum of open string ground states is labelled by $\sutwo$ representations $j'=0,1,...,j'_\max$, where $j'_\max = \min(2\jb',k-2\jb')$, and as usual $m'\in\{-j',...,j'\}$ (for a review, see for instance~\rcite{Schomerus:2002dc}).



\section{Symmetric products for open strings}
\label{sec:OpenSymProd}

\begin{figure}[ht]
\centering
    \includegraphics[width=.3\textwidth]{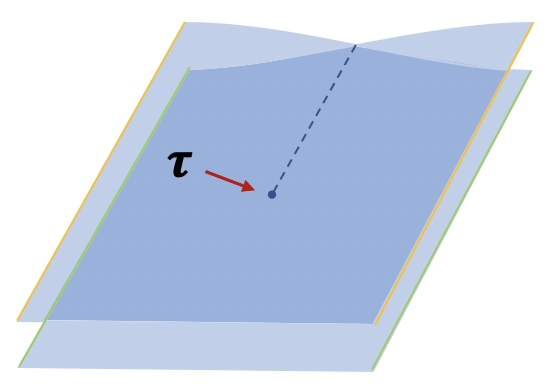}
\caption{\it 
Closed string joining/splitting is implemented by a $\bZ_2$ twist operator $\taub$ in the symmetric product CFT.  The orange (respectively green) edges are identified.  
}
\label{fig:closed-join}
\end{figure}

As shown in~\rcite{Balthazar:2021xeh}, type IIB $AdS_3$ string theory at $k<1$ is asymptotically free.  On the worldsheet, asymptotic scattering states consist of a Fock space of strings winding the $AdS_3$ azimuthal direction at asymptotically large radius $\rho$.  The spacetime CFT encodes this by having a symmetric product structure $(\cM)^p/\Gamma_p$, which describes a Fock space of free strings, with $\cM$ given in~\eqref{symprod}.  This symmetric product deformed by a marginal $\bZ_2$ twist operator that implements string joining/splitting interactions, see figure~\ref{fig:closed-join}.  The deformation has the schematic structure 
\eqn[closedtwist]{
\taub = e^{\beta\phi}\, \sigmab_2^{~}
}
where $\sigmab_2^{~}$ is a particular $\bZ_2$ twist operator that intertwines pairs of copies of $\cM$.  The radial profile $e^{\beta\phi}$ falls to zero in the asymptotic region $\phi\to\infty$ so that the theory is {\it asymptotically free}.  

In the open string sector, we have two classes of long open strings, protruding on either side of the $AdS_2$ brane.  The open string endpoints source equal and opposite charges for the $U(1)$ gauge theory on the brane; we can characterize a long open string by which boundary of $AdS_2$ the positively charged end asymptotes to in the far past, for instance (note that this end asymptotes to the opposite boundary in the far future, see figure~\ref{fig:LongOpen}).%
\footnote{In~\rcite{Lee:2001xe}, a second class of long open strings was proposed in which the endpoints asymptote to the same side in the far past and far future; however, further inspection reveals that these solutions do not have their endpoints on the same $AdS_2$ brane, but instead end on two different $AdS_2$ branes with conjugacy classes $\mu$ and $-\mu$.}

The worldsheet theory has interactions in which two oppositely charged string endpoints meet and join into a single connected string.
The boundary condition that describes the string after such a transition was described in~\rcite{Bachas:2001vj}, where a variety of gluing conditions of tensor product CFT's was described.  In addition to factorized boundary conditions where the chiral algebra obeys reflecting boundary conditions in the separate tensor factors, one can have perfectly transmissive boundary conditions wherein the left-movers of one tensor factor $\cM^+$ reflect into the right-movers of another tensor factor $\cM^-$ and vice-versa.  One can then ``unfold'' the two parameter space strips on which the open string theory is defined onto a single strip twice as wide (see figure~\ref{fig:unfold}), with the chiral algebra transmitting seamlessly between the domains describing $\cM^+$ and $\cM^-$.  In the present context, such a boundary condition describes a string worldsheet that, rather than ending on the D-brane, instead passes right through it without noticing its presence.

On this larger strip, one can again have ``untwisted'' (\ie\ reflecting) boundary conditions, or boundary conditions that intertwine additional copies of $\cM^\pm$.  The two kinds of tensor factor, $\cM^+$ and $\cM^-$, describe long open strings of unit winding that extend on the convex (for $\cM^+$) or on the concave (for $\cM^-$) side of the $AdS_2$ brane in the far past (equivalently, they are characterized by whether the positively charged end of the string approaches the left or the right boundary of the $AdS_2$ D-brane worldvolume in the far past).  The twisted boundary conditions only sew together $\cM^+$ to $\cM^-$.  The untwisted boundary condition describes strings that end on the D-brane.

Thus the spacetime CFT mirrors its worldsheet description discussed at the end of the previous section, where the brane is a defect in the conformal field theory on the sphere (in Euclidean $AdS_3$; the infinite cylinder for Lorentzian $AdS_3$).  We fold the coordinate domain of one side of the defect onto the other side, and describe the spacetime CFT on the upper half plane (Euclidean) or strip (Lorentzian) as a BCFT.  Strings that end on the defect are described by reflecting boundary conditions, while strings that pass through the defect are described by twisted boundary conditions that intertwine tensor factors $\cM^+\leftrightarrow\cM^-$.

If we interleave $w$ copies $\cM^\pm\to\cM^\mp\to\cM^\pm\cdots$ in this way, we generically have a long open string winding $w/2$ times around $AdS_3$; if $w$ is even and the intertwinings close into a cyclic permutation, one has a closed winding string described in open string language.

An explicit representation of the boundary conditions is given as follows.  Consider a set of $2p$ letters $\{1,\bar1,2,\bar2,\dots,p,\bar p\}$ representing the left- and right-movers for each copy of $\cM^\pm$.  The first $2p_+$ entries represent the left- and right-movers of the copies of $\cM^+$ while the remaining $2p_-$ entries represent the left- and right-movers of the copies of $\cM^-$.  The boundary conditions themselves are $2p\times 2p$ matrices specifying which left-mover reflects into which right-mover.  The standard boundary condition which reflects left- and right-movers within the $N^\th$ factor,%
\footnote{Our notation is that $N=1...p$, with $N=a=1...p_+$ labelling the copies of $\cM^+$, and $N-p_+=j=1...p_-$ denoting the $p_-$ copies of $\cM^-$.}
$\cM^\pm_N$, is the matrix $\sigma_1$ acting within the $2\times2$ diagonal block in positions $(2N-1,2N)$, $N=1,...,p$; thus the untwisted sector is represented by the matrix 
\eqn[gam0]{
\gammab_0 = \left( 
\begin{array}{ccc|ccc}
\sigone & ~ & ~ & ~ & ~ & ~ \\
~ & \ddots & ~ & ~ & ~ & ~ \\
~ & ~ & \sigone & ~ & ~ & ~ \\[.1cm]
\hline
~ & ~ & ~ & \sigone\rule{0pt}{2.8ex} & ~ & ~ \\
~ & ~ & ~ & ~ & \ddots & ~ \\
~ & ~ & ~ &  ~ & ~ & \sigone
\end{array}\right)  ~,
} 
Twisted boundary conditions consist of replacing disjoint pairs $(a_\alpha,j_\alpha)$ of diagonal $2\times2$ blocks in $\gammab_0$ by off-diagonal blocks that reflect the $(L,R)$ currents of the $a^\th$ copy  of $\cM^+$ into the $(R,L)$ currents of the $j^\th$ copy of $\cM^-$
\footnote{Abstractly, there are more general boundary conditions, for instance that reflect \eg\ $(1\to \bar 2, 2\to \bar 3,3\to\bar 1; \bar 1\to 3,\bar2\to1,\bar3\to2)$, so that left- and right-movers of a given string reflect into right- and left-movers of different strings, respectively.  Such boundary conditions generate the structure of a three-string (or in general multi-string) junction, rather than something related to perturbative string theory.  We will not consider them further in this work (though the general symmetric product boundary states considered in~\rcite{Belin:2021nck,Gaberdiel:2021kkp} have this character).}
\eqn[gamtranspose]{
\gammab^{(aj)} = \left( 
\begin{array}{ccc|ccc}
\ddots~ & ~ & ~ & ~ & ~ & ~ \\
~ & 0 & ~ & ~ & \sigone & ~ \\
~ & ~ & \ddots~ & ~ & ~ & ~ \\
\hline
~ & ~ & ~ & ~\ddots & ~ & ~ \\
~ & \sigone & ~ & ~ & 0 & ~ \\
~ & ~ & ~ &  ~ & ~ & ~\ddots
\end{array}\right)  ~;
} 
one can replace up to ${\textit{min}}(p_-,p_+)$ such block pairs in $\gammab_0$ in this manner to arrive at the generic twisted boundary condition.  One then must symmetrize separately over the label sets $a=1...p_+$, $j=1...p_-$ to project onto physical states.
%

The generic asymptotic state of multiple long open and closed strings is described by a pair of such boundary condition matrices $\gammab_L,\gammab_R$ specifying the boundary conditions for the currents on the left and right ends of the strip parametrizing the Lorentzian BCFT  
\eqn[JBC]{
J_M = \gammab^{(MN)} \, \bar J_N  
} 
consisting of some number of reflecting boundary conditions~\eqref{gam0} and some number of twisted boundary conditions~\eqref{gamtranspose}.
As mentioned above, closed string sectors are embedded in the twisted sectors of the open string theory.  The pair $(\gammab_L,\gammab_R)$ threads the left- and right-moving currents through the various copies; long open strings terminate in a reflecting boundary condition, while long closed strings consist of a sequence of intertwinings $\cM^+\leftrightarrow\cM^-$ that form a closed cycle. 

Note that the boundary condition matrices $\gammab_L$ and $\gammab_R$ on the left and right edges of the strip are not themselves permutations, but rather intertwiners between left- and right-movers.  Their product $\pib=\gammab_L(\gammab_R)^T$, though, factorizes into a permutation of the left-movers of the various string strands among themselves, and an identical permutation of the right-movers among themselves.  The letters in the permutation can be bicolored, say blue for the copies of $\cM^+$ and red for the copies of $\cM^-$.  As an example, suppose we have $p_+=5$, $p_-=3$, and 
\eqna[gammas]{
\gammab_L &= (\bl1) (\bl2\,\rd6) (\bl3) (\bl4\,\rd8) (\bl5\,\rd7)
\\[.2cm]
\gammab_R &= (\bl1) (\bl2) (\bl3\,\rd6) (\bl4\,\rd7) (\bl5\,\rd8)
\\[.2cm]
\pib &= (\bl1) (\bl2\,\bl3\,\rd6) (\bl4\,\bl5) (\rd8\,\rd7) ~.
}
That is, on the left edge we have reflecting boundary conditions on strands 1 and 3, and intertwining boundary conditions reflecting copy 2 of $\cM^+$ into the first copy of $\cM^-$ (or rather $N=6$, corresponding to $j=N-p_+=1$); similarly $N=4,8$ and $N=5,7$ are intertwined; and so on for the right movers.
All told, the boundary conditions describe a single untwisted open string (strand 1), a length three open string (strands 2-6-3) making three half-turns around AdS, and a winding two closed string (strands 4-8-5-7).

The permutation $\pib$ consists of a product of disjoint cycles.  If a cycle in $\pib$ contains both colors, it describes an open string, while if it contains only a single color, it describes half the open string strands that are sewn into a closed string, and there is another cycle of the same length having only the complementary color. 

The collection of cycles in $\pib$ is insufficient to determine $\gammab_L$ and $\gammab_R$ separately without some further conventions, as they doesn't specify on which copies of $\cM^\pm$ an open string begins and ends; and it doesn't specify the the way the two separate cycles describing a closed string interleave.  

Let us specify some additional conventions so that the transpositions in $\gammab_L,\gammab_R$ can then be read off from $\pib$ as follows.  Suppose a cycle in $\pib$ is bicolored, for example
\eqna[pidecoding1]{
\pib &\supset \big(\bl{a_1\,a_2\dots a_n}\,\rd{j_1\,j_2\dots j_{n}}\big)
\longrightarrow 
\begin{cases}
\gammab_L \supset (\bl{a_1}\,\rd{j_1})(\bl{a_2}\,\rd{j_2})\cdots\cdots (\bl{a_n}\,\rd{j_n}) \\
\gammab_R \supset (\bl{a_1})(\rd{j_1}\,\bl{a_2})\cdots (\rd{j_{n-1}}\,\bl{a_{n}})(\rd{j_n})
\end{cases} ~;
}
then one has an open string.
The endpoints of the string are charged; we take the right end of $\cM^+$ and the left end of $\cM^-$ to be negatively charged, and adopt a convention that the cycle begins at the negatively charged end of the string which we take to be the first letter in the cycle.  Thus the above cycle corresponds to an open string which begins at the right end of $\cM^+_{a_1}$ and ends at the right end of $\cM^-_{j_n}$
If the number of red and blue letters in the cycle differ by one, then the begins on one side of the strip and ends on the other, for instance
\eqna[pidecoding2]{
\pib &\supset \big(\rd{j_1\,j_2\dots j_{n}}\,\bl{a_1\,a_2\dots a_{n-1}}\big)
\longrightarrow 
\begin{cases}
\gammab_L \supset (\rd{j_1})(\bl{a_1}\,\rd{j_2})~\cdots ~(\bl{a_{n-1}}\,\rd{j_n}) \\
\gammab_R \supset (\rd{j_1}\,\bl{a_1})\cdots (\rd{j_{n-1}}\,\bl{a_{n-1}})(\rd{j_n})
\end{cases} ~.
}

Closed strings consist of a pair of cycles, one red and one blue, which we place adjacent to one another in $\pib$.
One reads off the left and right intertwiners as
\eqna[pidecoding3]{
\pib &\supset \big(\bl{a_1\,a_2\dots a_n}\big)\big(\rd{j_1\,j_2\dots j_{n}}\big)
\longrightarrow 
\begin{cases}
\gammab_L \supset (\bl{a_1}\,\rd{j_1})(\bl{a_2}\,\rd{j_2})\cdots\cdots (\bl{a_n}\,\rd{j_n}) \\
\gammab_R \supset (\rd{j_1}\,\bl{a_2})\cdots (\rd{j_{n-1}}\,\bl{a_{n}})(\rd{j_{n}}\,\bl{a_{1}})
\end{cases}
}

The boundary conditions are thus labelled by a bicolored permutation.  There is a natural action of $S_{p_+}\times S_{p_-}$ which permutes the letters while preserving the bicoloring.  Physical string states are obtained by symmetrizing over this group, which are thus labelled by a decorated conjugacy class of $S_p$, where the decoration consists of a specification of a bicoloring for each cycle.  Closed strings correspond to pairs of equal length cycles, one of which is red and the other blue.  Open strings correspond to a single bicolored cycle together with a choice of which color comes first; for odd length cycles, that color has an extra letter as we saw in~\eqref{pidecoding2}.

Open string vertex operators are intertwiners between a pair of boundary conditions $(\gammab,\gammab')$ on one edge of the string worldsheet, and so are labelled by the same data as the long string states.


\section{Deformed boundary conditions in the spacetime CFT}
\label{sec:bdydef}

The symmetric product structure~\eqref{opensymprod} thus describes the Fock space of long open strings.  In addition to the standard closed string joining/splitting interaction, there are boundary interactions where a long string breaks apart at a point where it intersects the brane, or where open string endpoints join up.  In this section we describe this boundary interaction.  As a warmup, let us review the structure of long string bulk interactions uncovered in~\rcite{Balthazar:2021xeh}.


\subsection{Review of the closed string deformation}
\label{sec:cldef}

An important role in the analysis of~\rcite{Balthazar:2021xeh} was played by non-normalizable operators associated to the analytic continuation of the continuous series, for instance the identity operators
\eqna[WSbulkidents]{
\cI^{(0)} &= e^{-\varphi-\bar\varphi}e^{iH_\sl+i\bar H_\sl}\,\Phi_{j=1;m=\mbar=-1}^{(w=0)}
\\[.2cm]
\cI^{(-1)} &= e^{-\varphi-\bar\varphi}e^{iH_\sl+i\bar H_\sl}\,\Phi_{j=1-\frac{k}{2};m=\mbar=\frac{k}{2}}^{(w=-1)} ~.
}
The first of these can be thought of as the zero mode of the dilaton; it also plays the role of the identity operator in the worldsheet realization of current algebra symmetries of the spacetime CFT~\rcite{Kutasov:1999xu}, and measures the number of wound fundamental strings dissolved in the fivebrane background.
In the regime $k>1$, the second operator $\cI^{(-1)}$ is (after the reflection $j\to 1-j$) the FZZ dual of $\cI^{(0)}$ and is thus a different representative of the same operator; while for $k<1$, it is the worldsheet representative of the identity operator in the in the block of the (deformed) symmetric product spacetime CFT (which in general maps to the $w=-1$ sector of the worldsheet theory~\rcite{Balthazar:2021xeh}).
Indeed, this operator is the special case $j'=m'=\mbar'=p_Y=\bar p_Y=0$ in~\eqref{blocktach}, which specializes the operator~\eqref{blocktachST} to $e^{\beta\phi}$; we have the correspondence  
\eqn[exptlop]{
e^{\beta\phi} ~\longleftrightarrow~
e^{-\varphi-\bar\varphi}e^{iH_\sl+i\bar H_\sl}\,\Phi_{j;m,\mbar}^{(w=-1)} ~,
}
with the quantum numbers, $\beta,j$ related by
\eqn[dimexp]{h_{\ST}=\bar h_{\ST} = -\half \beta(\beta+Q_\ell)= -\frac{j(j-1)}{k}+\frac{k-2}{4} ~.}
In the worldsheet theory, the mass shell condition determines $m,\bar{m}$ to be
\eqn[massshell]{m=\bar m=\frac{j(j-1)}{k}+\frac{k+2}{4}~.}
Setting in addition $\beta=0$ due to~\eqref{tachshell}, \eqref{expII}, one arrives at the identity operator in the block theory $\cM$.
For additional field content, notations and conventions see Appendix~\ref{sec:conventions}.

This operator has spacetime scaling dimension $(0,0)$, and satisfies 
\eqn[derzero]{\partial_x\cI^{(-1)}(x,\bar x)=0=\partial_{\bar x}\cI^{(-1)}(x,\bar x) ~.}
The worldsheet analysis of this result proceeds by evaluating 
\eqn[commJ]{\partial_x\cI^{(-1)} = [J_0^-,\cI^{(-1)}]=e^{-\varphi-\bar{\varphi}}e^{i \bar{H}_{\sl}}\left(\sqrt{2} \, \psi^3_{\sl} \,
\Phi^{(-1)}_{1-\frac{k}{2};\frac{k}{2},\frac{k}{2}}
+e^{iH_{\sl}} \, j^-_{\sl;0} \, \Phi^{(-1)}_{1-\frac{k}{2};\frac{k}{2},\frac{k}{2}}\right)  ~.}
and showing that it is BRST exact: 
\eqna[Lmbda]{\big\{Q_{\it BRST},e^{-2\varphi-\bar\varphi}\partial\xi e^{i\bar{H}_{\sl}}\Phi^{(-1)}_{j;m,\bar{m}}\big\}
&= e^{-\varphi-\bar\varphi} \,
e^{i \bar{H}_{\sl}}\left[\sqrt{2}{\left(\frac{k+2}{2}-m\right)} \, \psi^3_{\sl} \,
\Phi^{(-1)}_{j;m,\bar{m}}\right.\cr
\hskip 1.5cm 
+e^{iH_{\sl}} \, j^-_{\sl;0} \,
&\Phi^{(-1)}_{j;m,\bar{m}}+\left.(m\tight+j\tight-1)
\left(\partial e^{i H_{\sl}}+e^{iH_{\sl}}\partial\varphi \right) 
\Phi^{(-1)}_{j;m-1,\bar{m}}\right] ~.}
For $j=1-\frac{k}{2},\, m=\bar{m}=\frac{k}{2}$, \eqref{Lmbda} reduces to \eqref{commJ}. We conclude that $\partial_x\cI^{(-1)}(x,\bar x)$ is BRST exact and thus vanishes in correlation functions of BRST invariant operators, and thus has the characteristics of the identity operator in the spacetime CFT.

The result holds because the operators on the RHS of~\eqref{commJ} are not sitting on an LSZ pole in the limit that could cancel the vanishing of the coefficients.  The closed string reflection coefficient in the $m$-basis is given by the Fourier transform of the two-point function
\eqna[2ptFT]{
&\cR(j,m,\mbar) \propto \int\!d^2x\,  x^{j+m-1} \bar x^{j+\mbar-1} \,
\big\langle \Phi_j(x,\bar x) \Phi_j(x',\bar x') \big\rangle
\\[.2cm]
&\hskip 1cm
\propto \frac{1\tight-2j}{\pi}\int\!d^2x\,  x^{j+m-1} \bar x^{j+\mbar-1} |1-x|^{-4j}
= \frac{\Gamma(j\tight+m)\Gamma(j\tight-\mbar)\Gamma(2\tight-2j)}{\Gamma(m\tight-j\tight+1)\Gamma(-\mbar\tight-j\tight+1)\Gamma(2j)} ~,
}
where in the correlator on the RHS we have extracted the power law scaling contribution rather than the contact term (the latter represents the direct amplitude while the former gives the reflected amplitude).
The reflection coefficients $\cR(j,m,\mbar)$ and $\cR(j,m-1,\mbar)$ are both regular in the limit $\beta\to0$ of the exponential operators $e^{\beta\phi}$~\eqref{blocktachST}, since neither corresponds to a state in the discrete series.  In the absence of such a singularity the RHS of~\eqref{commJ} vanishes in the limit and one arrives at the first equality of~\eqref{derzero}.

A similar analysis, with left and right-movers exchanged on the worldsheet and in spacetime, shows that $\partial_{\bar x}\cI^{(-1)}(x,\bar x)$ vanishes. Thus, the operator $\cI^{(-1)}(x,\bar x)$ is a dimension zero operator, whose correlation functions are independent of position. It is natural to interpret it as the identity operator in the block of the symmetric product.\footnote{Though there are subtleties, see~\rcite{Giveon:2001up,Porrati:2015eha}.  More precisely, the coupling to $\cI$ on the worldsheet is the chemical potential for fundamental strings in a grand canonical ensemble.}

More generally, one didn't have to immediately take the limit $\beta\to0$ of the exponential operators $e^{\beta\phi}$. 
One has the correspondence
\eqna[dualbb]{\beta\partial_x\phi \, e^{\beta\phi} ~~\longleftrightarrow~~ 
(m+j-1)(\partial\varphi+i\partial& H_\sl)e^{-\varphi-\bar\varphi} \, e^{i(H_\sl+\bar H_\sl)} \, \Phi_{j;m-1,\bar m}^{(-1)}\cr
\hskip 1.5cm 
&{-\sqrt{2}\left(m-\frac{k}{2}\right)e^{-\varphi-\bar\varphi}\psi^3_\sl e^{i\bar{H}_\sl}\Phi_{j;m,\bar m}^{(-1)} ~.}
}
One can similarly take the derivative $\partial_{\bar x}$, and then carefully take the limit $\beta\to 0$.  The leading term on the LHS vanishes linearly in $\beta$, while naively the terms on the RHS vanish as $\beta^2$, tempting one to conclude that $\partial_{\bar x}(\partial\phi)=0$.  However, the term involving $\Phi^{(-1)}_{1-\frac k2;\frac k2-1,\frac k2-1}$ sits on an LSZ pole in the limit (see the discussion in the Appendix around equation~\eqref{modphiop}) and so also only vanishes linearly in $\beta$.
One has
\eqna[holoviol]{
&
\lim_{j\to 1-\frac k2,m\to\frac k2}
\left[\bar{J}_0^-,(\partial\varphi+i\partial H_\sl)e^{-\varphi-\bar\varphi} \, e^{i(H_\sl+\bar H_\sl)} \, 
\Phi^{(-1)}_{j;m-1,\bar{m}}\right]
\\[.2cm]
& \hskip 2cm
= e^{-\varphi-\bar\varphi}\,
e^{i(H_\sl+\bar H_\sl)}\, 
(\partial\varphi+i\partial H_\sl) 
(\bar\partial\bar\varphi+i\bar\partial \bar{H}_\sl)\,
\Phi^{(-1)}_{1-\frac k2;\frac k2-1,\frac k2-1}~. 
}
The operator on the RHS is normalizable, and so has an FZZ dual in winding two
\eqn[weq2wall]{
e^{-\varphi-\bar\varphi}\,
e^{i(H_\sl+\bar H_\sl)}\, 
(\partial\varphi+i\partial H_\sl) 
(\bar\partial\varphi+i\bar\partial \bar{H}_\sl)\,
\Phi^{(-2)}_{k;k,k}  
}
whose dual in the spacetime CFT is a $\bZ_2$ twist operator $\taub$ that implements string joining/splitting interactions.  This operator is the sum of the top components of the $\half$-BPS twist operators of the symmetric product CFT 
\eqn[twist2]{
\sigmab^\pm = \sigmab^\pm_{\rm free}\,\sigmab^\pm_\LG = \exp\Big[ -\frac{1}{2\sqrt{k}}\big( \phi_S \mp iY_S\big) \Big]\,
\big( \sigmab_{\phi^{~}_A} \sigmab^{~}_{Y_A} \sigmab^{\pm}_{\chi_A} \big) \, \sigmab^\pm_{\LG}  
}
where we have suppressed the analogous antiholomorphic structure, whose chiral/antichiral nature (the $\pm$ label) can be chosen independently.
The conclusion of the analysis of~\rcite{Balthazar:2021xeh} is that this operator is turned on in the background, so that the non-holomorphy of $\partial\phi$ agrees between worldsheet and spacetime; in addition, this operator implements interactions of the long strings in the spacetime CFT.  

The deformation of the symmetric product $(\cM)^p/S_p$ for long closed strings is thus arrived at via the consideration of holomorphic operators in the block theory $\cM=\bR_\phi\times\bS^3_\flat$ of the symmetric product; the deformation softly and spontaneously breaks their holomorphy.

Note that the $\phi$ dependence is such that the interaction turns off in the asymptotic region $\phi\to\infty$, justifying the symmetric product starting point.  The coefficient of this deformation can be absorbed in a shift of $\phi$, and so there is no free parameter in the theory, just a scale in $\phi$ where the interactions become of order one.  One thus has a holographic version of the asymptotic freedom familiar from gauge theory.


\subsection{Open string version}

The rationale for the identification of $\cI^{(-1)}$ with the identity operator in the block theory $\cM$ of the spacetime symmetric product CFT $(\cM)^p/S_p$ comes from the fact that it has spacetime dimension $h=\bar h=0$ and is translation invariant
in the worldsheet representation of spacetime CFT correlation functions.

Repeating this analysis for boundary operators, one has the correspondence
\eqna[bdyexptl]{
e^{\beta\phi/2} ~~\longleftrightarrow~~ 
& e^{-\varphi/2}\,\psi^+_\sl\,\Psi_{j;m}^{(-1)} ~.
}
The worldsheet boundary operators
\eqna[WSbdyidents]{
\cI^{(0)}_\partial &= e^{-\varphi/2}\,\psi^+_\sl\,\Psi_{j=1;m=-1}^{(w=0)}
\\[.2cm]
\cI^{(-1)}_\partial &= e^{-\varphi/2}\,\psi^+_\sl\,\Psi_{j=1-\frac{k}{2};m=\frac{k}{2}}^{(w=-1)}
}
are the analogous candidates for worldsheet representatives of the boundary identity operator.%
\footnote{Note the halving of free field exponents such as that of $\varphi$ due to~\eqref{bdydim}.}
The first of these is associated with the zero mode of the open string gauge field; indeed, because the endpoints of open string are charged under this gauge field, open strings dissolved in an $AdS_2$ brane generate a constant electric field on the brane.  It also plays the role of the identity operator in the algebra of {\it spacetime} CFT boundary operators~\rcite{Giveon:2001uq}.  The second operator $\cI^{(-1)}_\partial$ is, for $k>1$ (and after the reflection $j\to 1-j$), the FZZ dual of $\cI^{(0)}_\partial$~\rcite{Kutasov:2005rr}; while for $k<1$, it is the obvious candidate for the boundary identity operator in a (deformed) symmetric product spacetime CFT.

The calculation of the boundary derivative of this candidate operator is completely analogous, with the result
\eqna[dbdyexptl]{
\partial_x \, e^{\beta\phi/2} ~~\longleftrightarrow~~ 
& \big[J^-_0 \; , \; e^{-\varphi/2}
\,\psi^+_\sl 
\,\Psi_{j;m}^{(-1)} \big]
\\[.2cm]
&\hskip1cm
= (m+j-1)(\partial\varphi\,\psi^+_\sl+\partial\psi^+_\sl)\,e^{-\varphi/2} \,  \Psi_{j;m-1}^{(-1)}\\[.1cm]
&\hskip 2cm
{-\sqrt{2}\left(m-\frac{k}{2}\right)e^{-\varphi/2}\,\psi^3_\sl \,\Psi_{j;m}^{(-1)} }
}
(again up to BRST exact quantities).  Unlike the closed string case, the first term on the RHS is sitting on an LSZ pole.  The open string reflection coefficient is proportional~to
\eqna[bdy2ptFT]{
\cR_\partial^{~}(j,m) &\propto \int_{-\infty}^\infty\!du\,  |u|^{j+m-1}  \,
\big\langle \Psi_j(u) \Psi_j(u') \big\rangle
\\[.2cm]
&\propto \int_{-\infty}^\infty\!du\,  |u|^{j+m-1}  |1-u|^{-2j}
\\[.2cm]
&= \frac{\Gamma(j+m)\Gamma(1-2j)}{\Gamma(-j+m+1)}
+ \frac{\Gamma(j-m)\Gamma(1-2j)}{\Gamma(-j-m+1)}
+ \frac{\Gamma(j+m)\Gamma(j-m)}{\Gamma(2j)}
\\[.2cm]
&= \frac{\Gamma(j+m)\Gamma(j-m)}{\Gamma(2j)} \;
\frac{\cos[\coeff\pi2(j+m)]\cos[\coeff\pi2(j-m)]}{\cos(j\pi)}
~.
}
Now the limit $\beta\to0$ 
{\it does} sit on an LSZ pole, and the RHS of the derivative of the candidate boundary identity operator $\cI^{(-1)}_\partial$ does not vanish, rather
\eqn[bdyidentderiv]{
\partial_x \cI^{(-1)}_\partial \sim
(\partial\varphi\,\psi^+_\sl+\partial\psi^+_\sl)\,
e^{-\varphi/2}\,\Psi^{(-1)}_{1-\frac k2;\frac k2-1}
~\longleftrightarrow~
\partial_x e^{-Q_\ell\phi/2}  ~.
}

Thus the candidate boundary ``identity operator'' in the winding one sector, $\cI^{(-1)}_\partial$, is not really the identity operator~-- rather it is a linear combination of the identity operator and the nontrivial dimension zero operator $e^{-Q_\ell\phi/2}$.  Nevertheless, it motivates a candidate for the worldsheet representative of the boundary twist interaction as follows.

As in the closed string case, the operator on the RHS of the worldsheet side of the correspondence~\eqref{bdyidentderiv} is the highest component of a spacetime chiral superfield. 
As is usual in free field representations of boundary operators, the boundary conditions can be accommodated by a suitable analytic extension of worldsheet fields across the boundary, so that the vertex operators look like the holomorphic part of closed string (``bulk'') vertex operators apart from a halving of the exponents as we saw in the discussion around~\eqref{bdydim}.  Reading this off from the analysis of~\rcite{Balthazar:2021xeh}, we have the operator
\eqn[Sbdy]{
\cS^\pm e^{-Q_\ell\phi/2} 
~~\longleftrightarrow~~
e^{-\frac\varphi4}\, e^{\frac{3i}4 H_\sl}\, 
e^{\pm\frac i4 (H_3+aZ+\sqrt{\frac2k}\,Y) } \,
\Phi^{(-1)}_{1-\frac k2;\frac k2-1} ~,}
where again $H_\sl,H_3$ bosonize $\psi^\pm_\sl,\psi^3_\su\tight\pm\psi^3_\sl$ respectively (see the Appendix), and $a^2=1-2/n$.
This operator is a marginal $\hf$-BPS operator in spacetime.

Acting with the spacetime supersymmetry generators $\cG^\pm_{-\half}=\oint S^\pm_{-\half}$ (where $S^\pm_r$ is given in~\eqref{susyops}), one finds the highest component as in~\rcite{Balthazar:2021xeh}
\eqna[NSwall]{
\big\{ \cG^\mp_{-\half} &, \, \cS^\pm  e^{-Q_\ell\phi/2} \big\}  
~\longleftrightarrow~ \cV^\pm
\\[.2cm]
&\cV^\pm \equiv 
-\frac 12\Big[ \partial\varphi+i\partial H_\sl \pm \Big(i\partial H_3+ ia\,\partial Z +i\sqrt{\frac2k}\,\partial Y\Big)\Big] \,
e^{-\varphi/4}\, e^{\frac i4 H_\sl}\, \Phi^{(-1)}_{1-\frac k2;\frac k2-1} 
~.
}
The sum of these two $\hf$-BPS operators $\cV^+ + \cV^-$
gives~\eqref{bdyidentderiv}.

As a perturbation of the block theory, \eqref{bdyidentderiv} is redundant, however
FZZ duality relates this normalizable worldsheet operator to a winding two operator (also normalizable)
\eqn[dualbdyderiv]{
(\partial\varphi\,\psi^+_\sl+\partial\psi^+_\sl)\,
e^{-\varphi/2}\,\Psi^{(-2)}_{k;k} ~.
}
According to our identification of boundary winding sectors, $w=-2$ corresponds to a single unit of string winding around the $AdS_3$ azimuthal direction.  Once again, we interpret this operator as the marginal twist deformation that implements string joining/splitting transitions, in this case sewing together the endpoints of open strings, or breaking a brane-string intersection into a pair of string endpoints.

\begin{figure}[ht]
\centering
    \includegraphics[width=.7\textwidth]{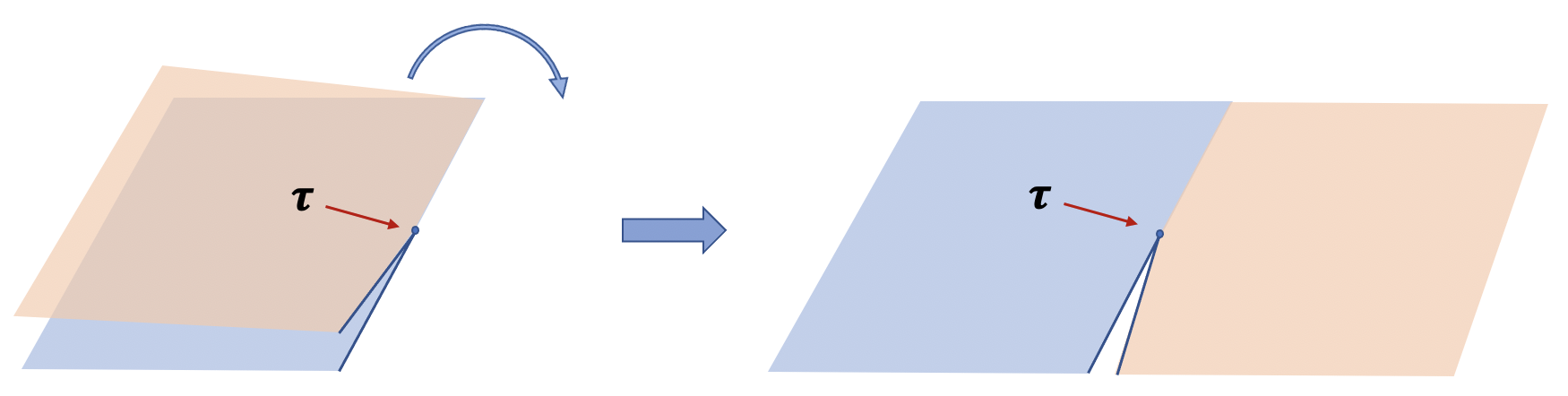}
\caption{\it 
Unfolding an open string joining/splitting operator $\taub$ that sews together copies of $\cM^+$ and $\cM^-$ to make an open string that winds once around $AdS_3$.  Dark edge lines indicate the locus of reflecting boundary conditions, while the absence of such a dark line at the boundary between $\cM^+$ (blue) and $\cM^-$ (red) indicates the locus of transparent boundary conditions.  
}
\label{fig:unfold-join}
\end{figure}


\subsection{The boundary twist operator}

We expect the operator corresponding to~\eqref{dualbdyderiv} in the spacetime CFT to be the boundary version of the bulk $\bZ_2$ twist operator that implements long string interactions in the bulk of its worldsheet, \ie\ it should implement the boundary splitting/joining transitions that generate the twisted sectors of the symmetric product discussed in section~\ref{sec:OpenSymProd}.

\begin{figure}[ht]
\centering
    \includegraphics[width=.9\textwidth]{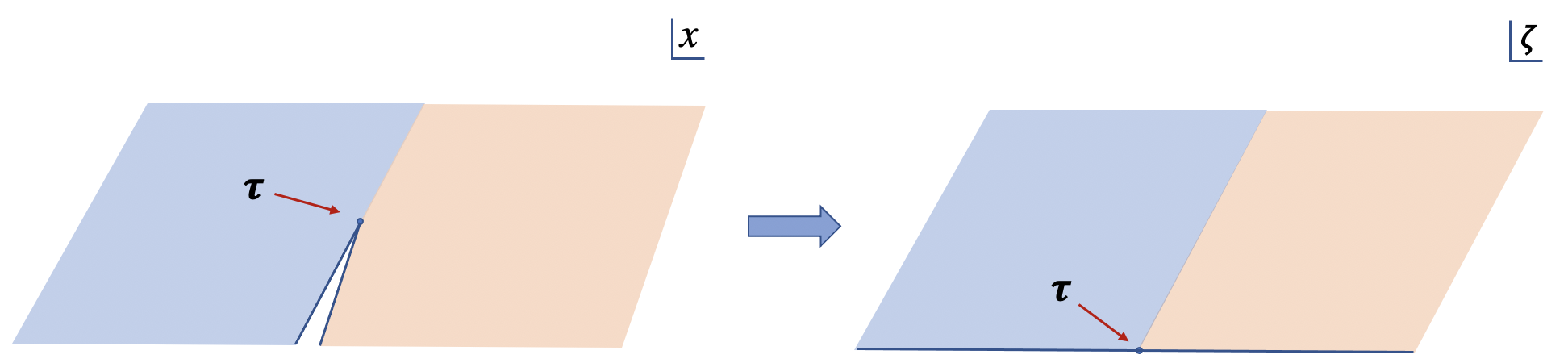}
\caption{\it 
A Schwarz-Christoffel transformation maps the open string joining/splitting interaction to the upper half plane.
}
\label{fig:SCtransf}
\end{figure}

In the symmetric product analysis of section~\ref{sec:OpenSymProd}, such an interaction arises when a copy of $\cM^+$ and a copy of $\cM^-$ are sewn together, see figure~\ref{fig:unfold-join}.  Unfolding the interaction, one finds a coordinate domain that can be mapped to the upper half plane by a Schwartz-Christoffel transformation
\eqn[SCtransf]{
\zeta \sim \sqrt{x} + {\it regular} ~.
}
This is the same sort of covering map that one uses for the closed string $\bZ_2$ twist interaction~\rcite{Lunin:2000yv}.  There it maps the ``pair-of-pants'' interaction of closed strings (see figure~\ref{fig:closed-join}) to the covering sphere or complex plane; here it maps the analogous interaction of open strings to the upper half plane (or equivalently the Poincar\'e disk), see figure~\ref{fig:SCtransf}.
%

The covering map turns the ground state boundary twist operator on the $x$ plane into the identity operator in the $\zeta$ plane.  Thus the Schwarzian derivative of this map determines the conformal dimension of the ground state boundary twist operator $\sigmab_\partial^{~}$ to be~\rcite{Dixon:1986qv}
\eqn[twistdim]{
h[\sigmab_\partial] = \frac{c}{16} = \frac{3k}{8} ~.
}
The $\half$-BPS twist deformation $\sigmab^\pm$ decorates the ground state twist operator with a center-of-mass exponential that carries the spacetime $R$-charge, as well as an exponential of $\phi$ to make the operator marginal and $\half$-BPS (for details, see Appendix~C of~\rcite{Balthazar:2021xeh})
\eqn[twist2open]{
\sigmab^\pm_\partial = \sigmab^\pm_{\rm free}\,\sigmab^\pm_\LG = \exp\Big[ -\frac{1}{4\sqrt{k}}\big( \phi_S \mp iY_S\big) \Big]\,
\big( \sigmab_{\phi^{~}_A} \sigmab^{~}_{Y_A} \sigmab^{\pm}_{\chi_A} \big) \, \sigmab^\pm_{\LG}  ~.
}
Here, we have decomposed the squashed $\sutwo$ into its $\uone$ and $\frac\sutwo\uone$ (\ie\ $N=2$ Landau-Ginsburg) components, and associated the $\uone$ factor with $\bR_\phi$ to make a free $N=2$ superfield.
The center-of-mass contributions of the fermions and Landau-Ginsburg fields are left implicit, with the superscript $\pm$ indicating whether these contributions make the operator chiral or antichiral.  The descendant under the supercurrent makes the twist two contribution to the marginal deformation
\eqn[marginaltwist]{
\taub_\partial^{~} = \cG^+_{-\half}\, \sigmab_\partial^- + \cG^-_{-\half}\, \sigmab_\partial ~.
}
We thus have a candidate correspondence between the above twist two boundary operator in the spacetime CFT, and the worldsheet winding two open string vertex operator, given by the upper component of~\eqref{dualbdyderiv} under the spacetime supersymmetry~\eqref{susyops}.

\begin{figure}[ht]
\centering
    \includegraphics[width=.5\textwidth]{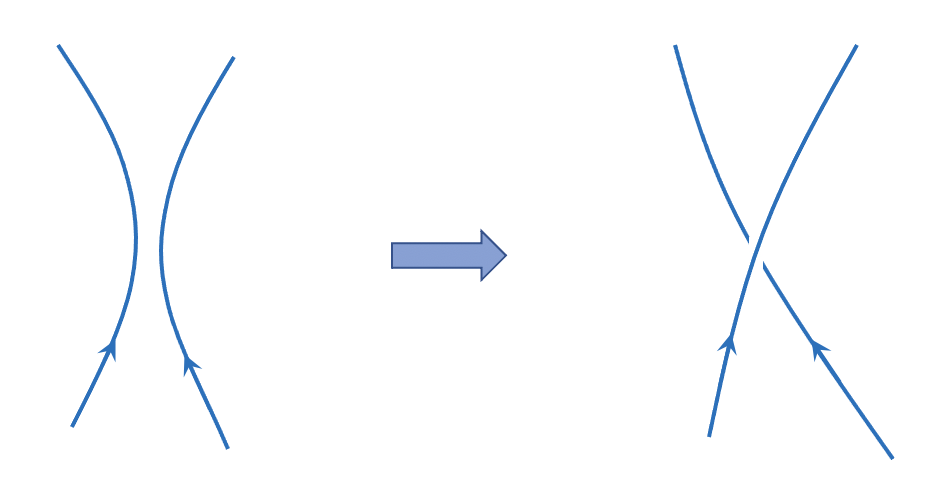}
\caption{\it 
A bulk twist operator exchanges open string segments and implements a $2\to2$ interaction of open strings as well as closed string joining/splitting, and open+closed~$\leftrightarrow$~open.
}
\label{fig:exchange}
\end{figure}

In addition to the boundary joining/splitting interaction, CFT blocks $\cM^+$ can interact among themselves pairwise via the bulk $\bZ_2$ twist interaction that implements an exchange of segments as depicted in figure~\ref{fig:exchange}.  Similarly, copies of $\cM^-$ can interact among themselves.
%
In principle, there should also be interactions in the bulk of the worldsheet in which $\cM^+$ exchanges segments with $\cM^-$.  These are not possible in the asymptotic region which the symmetric product describes, since $\cM^+$ and $\cM^-$ describe strings stretched on opposite sides of the $AdS_2$ brane, and such strings only meet on the brane.  
However, as we see from figure~\ref{fig:LongOpen}, such strings collapse toward the center of $AdS_3$ where they can meet and interact.  
In the BCFT~\eqref{opensymprod}, a direct interaction of $\cM^+$ with $\cM^-$ along the lines of figure~\ref{fig:exchange} would violate winding conservation by generating two strings, half coming from $\cM^+$ and half from $\cM^-$ (in particular, their ends approach the $AdS_2$ brane from opposite sides); thus one has two strings each winding once around $AdS_3$.  The possibility of such an interaction seems to be tied up with the way that the boundary CFT realizes the center of $AdS_3$ and the winding-violating processes that can occur there in the worldsheet theory.


\section{Discussion}
\label{sec:discussion}

The considerations above extend the $AdS/CFT$ dictionary for the stringy regime $k<1$ to a duality involving boundary conformal field theory.
The boundary CFT dual to an $AdS_2$ brane in $AdS_3$ takes the form of the symmetric product~\eqref{opensymprod}, deformed by the bulk $\bZ_2$ twist field (the top component of~\eqref{twist2}) that implements the bulk joining/splitting interaction of figures~\ref{fig:exchange}, together with the boundary joining/splitting interaction of figure~\ref{fig:unfold-join} given by~\eqref{marginaltwist}.

\vskip .8cm
\noindent
\ref{sec:discussion}.1~~{\em Relation between bulk and boundary deformations of the symmetric product}
\medskip

In the worldsheet theory, the strength of these two types of  strength of these two interactions is related by $g_\partial^2 = g_{\rm bulk}$; in the spacetime CFT they are a priori unrelated (as are the relative strengths of the separate $\bZ_2$ twist operators in each tensor factor of~\eqref{opensymprod}, for that matter).  

Further support for such a relation between the set of boundary and bulk interactions comes from their similarity to those of light cone gauge string field theory~\rcite{Kaku:1974zz}.  Indeed, there have been proposals to gauge fix $AdS_3$ string theory to a sort of light-cone gauge (or static gauge, which for long strings amounts to the same thing)~\rcite{Yu:1998qw,Hikida:2000ry}.  However, such a gauge on the worldsheet only makes sense asymptotically and in the long string sector, where one can work in a Wakimoto representation and perturb in the exponential interaction; in this approximation the worldsheet dynamics is free, and allows a light-cone gauge choice.  Nevertheless, perturbative amplitudes under such assumptions have much the same structure as light cone string field theory.  The Feynman diagrams of that field theory cover the moduli space of Riemann surfaces~\rcite{Giddings:1986bp}, but only yield a unitary amplitude if the open and closed string interactions are related by $g_{\rm open}^2=g_{\rm closed}$.  This suggests a similar relation in our boundary conformal field theory.  

\begin{figure}[ht]
\centering
    \includegraphics[width=.9\textwidth]{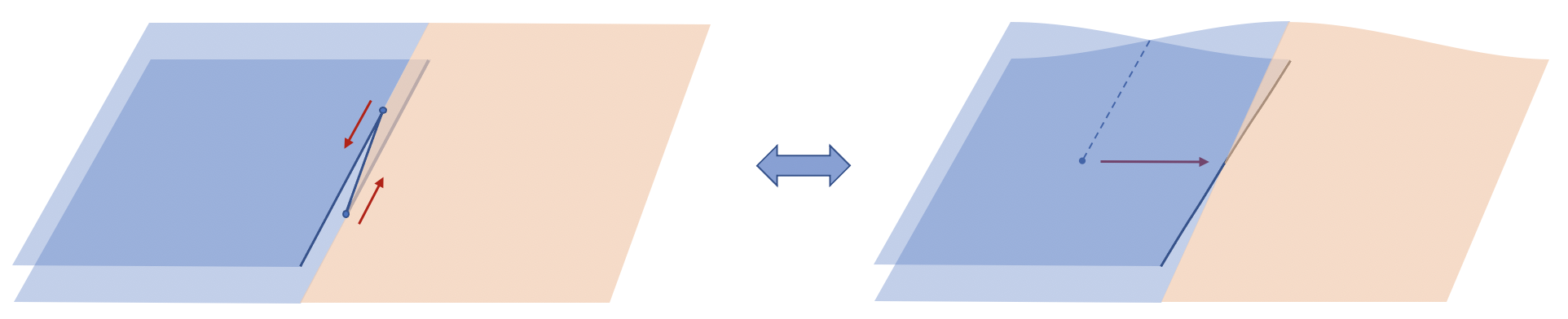}
\caption{\it 
The transition in which a pair of boundary intertwining operators collide and form a bulk $\bZ_2$ twist operator.  Again, dark edge lines indicate the locus of reflecting boundary conditions, while the absence of such a dark line at the boundary between $\cM^+$ (blue) and $\cM^-$ (red) indicates the locus of transparent boundary conditions.  The bulk interaction on the right-hand figure is a transition of the sort depicted in figures~\ref{fig:closed-join} and~\ref{fig:exchange}; the reflecting boundary condition lives entirely on one copy of $\cM^+$, while the other copy of $\cM^+$ transmits into the copy of $\cM^-$ all along its boundary.
}
\label{fig:opxop-to-cl}
\end{figure}

The boundary intertwining operator $\taub_\partial^{(aj)}(u)$ that sews together $\cM^+_a$ and $\cM^-_j$ and the boundary intertwining operator $\taub_\partial^{(aj')}(u')$ intertwining $\cM^+_a$ and $\cM^-_{j'}$ should have a coincidence limit $u'\to u$ that matches smoothly onto the boundary limit $x\to u$ of the bulk $\bZ_2$ twist operator $\taub^{(jj')}(x)$; see figure~\ref{fig:opxop-to-cl}.  
Such a relation would reproduce a corresponding property of light cone string field theory (see figures 10-13 of~\rcite{Kaku:1974zz}) that relates the open and closed string joining/splitting interactions (and in our case, the $2\to 2$ open string interactions on $\cM^+$ and on $\cM^-$).  The open and closed string vertices generate a diagrammatic expansion which realizes a cell decomposition of the moduli space~\rcite{Giddings:1986rf}.  The matching of the closed and open string amplitudes amounts to their proper factorization and is a necessary condition for various Ward identities to be satisfied. 
It would be useful to understand this point in detail.


\vskip .8cm
\noindent
\ref{sec:discussion}.2~~{\em The dual of the $AdS_2$ brane parameter $\mu$}
\medskip

Thw $AdS_2$ brane comes with a parameter $\mu$ specifying a worldvolume electric field created by dissolved fundamental strings.  Thus far, we have not specified the corresponding quantity in the spacetime BCFT.  

In the closed string theory, the winding charge $p$ is the Legendre transform of the the corresponding chemical potential, which is the coupling to the operator $\cI^{(-1)}$%
~\rcite{Giveon:1998ns,Kutasov:1999xu,Giveon:2001up,Porrati:2015eha}.
Similarly, the $AdS_3$ string coupling is the zero mode of the dilaton, which is the operator $\cI^{(0)}$.  In the critical dimension, these two are related by the fixed scalar condition $\gstrsq=kV_{\bT^4} /p$, as well as the fact that the two operators $\cI^{(0)}$, $\cI^{(-1)}$ are FZZ dual, \ie\ different representatives of a single operator.

The corresponding open string quantities are the operators $\cI^{(0)}_\partial$, the zero mode of the electric field on the brane, and $\cI^{(-1)}_\partial$, the FZZ dual condensate of stretched open strings dissolved in the brane that create this electric field.  The natural parameter related to this electric field is the asymmetry $\frac{p_+-p_-}{p_++p_-}$; after all, $p_+-p_-$ represents the net charge on one end of the brane and thus an asymptotic source of electric field on the brane~-- or rather the Legendre transform of it.  As in the closed string case where $p$ is conjugate to the bulk coupling to $\cI^{(-1)}$, here we expect that $p_+-p_-$ is conjugate to the boundary coupling to $\cI^{(-1)}_\partial$ (in both senses of duality -- AdS/CFT duality and Legendre duality between charges and field zero modes).

It might be interesting to understand the limit when there are only \eg\ $\cM^+$ strings, which would appear to correspond on the worldsheet to the limit of critical electric field on the brane.  In this limit, the electric field on the brane compensates the tension of the fundamental string, which becomes effectively tensionless.  Some aspects of this limit were studied in~\rcite{Lee:2001xe}.  In our candidate spacetime dual, as the asymmetry $\cA=\frac{p_+-p_-}{p_++p_-}$ increases in magnitude, there are fewer and fewer ways of twisting together blocks of the symmetric product~\eqref{opensymprod} to make higher winding strings, until at $|\cA|=1$ there can be no boundary twist interaction, leaving only the bulk $\bZ_2$ twist of figures~\ref{fig:closed-join} and~\ref{fig:exchange}.

While the bound state strings in the discrete series depicted in figure~\ref{fig:OpenStrings} keep to one side of the brane and contribute to a static electric field, scattering states reverse orientation during their evolution (see figure~\ref{fig:LongOpen}); their back-reaction reduces the field on the brane during the scattering process.  This fact seems somewhat in tension with the identification of the asymmetry $\cA$ with the electric field on the brane; on the other hand, the worldsheet theory operates in a perturbative regime where the number of scattering states is small relative to the number of strings dissolved in the brane background, so perhaps the fact that a macroscopic number of scattering states will reverse the sign of the brane electric field during the course of evolution is not is not in direct contradiction with the identification of $\cA$ with the electric field.


\vskip .8cm
\noindent
\ref{sec:discussion}.3~~{\em Defective fuzz}
\medskip

As $AdS_3$ becomes more stringy with decreasing radius of curvature $k$, D-brane wavefunctions become more delocalized.  This phenomenon was studied in the closely related $\frac\sltwo\uone$ coset model in~\rcite{Kutasov:2005rr}. In the coset $\frac\sltwo\uone$, the $AdS_2$ brane descends to the same one-dimensional worldvolume~\eqref{conjclasscoords} having the shape of a ``hairpin'' in the cigar-like coset geometry~\rcite{Quella:2002ns,Ribault:2003ss,Fotopoulos:2003vc,Israel:2005fn}.   The shape deformation corresponding to the conjugacy class parameter $\mu$ (the coefficent of the boundary operator $\cI^{(0)}_\partial$ of equation~\eqref{WSbdyidents}) descends to a parameter controlling the location of the tip of the hairpin relative to the tip of the cigar.  Its FZZ dual $\cI^{(-1)}_\partial$ (for $k>1$) represents a condensate of stretched open strings at the tip of the hairpin. 

It was argued in~\rcite{Kutasov:2005rr} that this open string condensate causes the wavefunction of the hairpin brane to delocalize in the cigar coset for $k<1$.  This result lifts directly to the properties of the $AdS_2$ brane in the parent $AdS_3$, since the vertex operator $\cI^{(-1)}_\partial$ is $\uone$ invariant (\ie\ involves no $Y$ exponential).  Thus, the $AdS_2$ brane appears to be delocalized in the $AdS_3$ throat.  The symmetric product partitions the coordinate domain of the CFT into two complementary domains (that we have painted red and blue in the analysis above), separated by a defect of fixed position.  The dual bulk description appears to be considerably fuzzier, with quantum fluctuations smearing the bulk brane all over $AdS_3$.  This feature appears correlated to the absence of a black hole horizon in highly excited states, in that both arise as a result of the correspondence transition discussed in~\rcite{Balthazar:2021xeh}.  In both cases, stringy fluctuations dominate the wavefunction and lead to a fuzzball in the bulk.

In the dual spacetime CFT, it is a curious feature of the deformed symmetric product that the two sides of the defect only interact at the defect, through the boundary intertwining operator.  The bulk dual of this is that, even though the $AdS_2$ brane is fluctuating wildly, it still divides $AdS_3$ into two disjoint bulk domains, albeit in a rather convoluted way.  Quantum entanglement of the degrees of freedom on either side occurs solely through this interaction.  Thus the deformed symmetric product structure of the spacetime CFT provides an interesting laboratory for the investigation of such entanglement structures and their relation to bulk geometry.  In quantum field theory, one can partition spacetime into disjoint regions, with the degrees of freedom entangled across domain boundaries in a unique way specified by the vacuum structure of the QFT (\ie\ every sensible state of the QFT has to asymptote to the same conformal fixed point in the UV, and so the UV entanglement structure across a domain boundary is universal).  It has always been a puzzle as to how such a partitioning might work in string theory, given that the UV structure of string wavefunctions respects no domain boundaries and is spread over all of spacetime~\rcite{Karliner:1988hd,Klebanov:1988ke}.%
\footnote{For instance, there is no sensible quantization of strings in a Rindler wedge (as far as the author is aware), because the string worldsheet can't be kept on one side of a null surface.}
Perhaps the present construction may offer some insight as to how partitioning the conformal boundary of $AdS$ extends into the bulk $AdS$ geometry, and thereby illuminate the structure of the holographic map.


\vskip .8cm
\noindent
\ref{sec:discussion}.4~~{\em Generalizations}
\medskip

There are a number of straightforward generalizations of our construction.  The generalization to $(\sfp,\sfq)$ strings was essentially treated above; the $\sfp$ D1's bind together and puff up into a dipolar D3 brane, as discussed in section~\ref{sec:classical}.  The resulting $AdS_2\times\bS^2$ brane carries the D1 charge $\sfp$ as a quantized worldvolume magnetic flux through $\bS^2$, and the F1 charge $\sfq$ as a worldvolume electric field along $AdS_2$.  There should also be $(\sfp,\sfq)$ string junctions, and string webs, which would chop up the boundary domain into multiple disjoint regions, generalizing the two sides of the $AdS_2$ brane.  In the limit of large $(\sfp,\sfq)$ the defect(s) should backreact classically, leading to various Janus-type solutions in regimes where supergravity is valid; and these should have counterparts in the stringy $AdS_3$ regime.

There are also other $\sltwo$ D-branes besides the $AdS_2$ brane associated to the conjugacy class~\eqref{SL2 conj class}.  In particular, branes associated to the conjugacy class of the identity in $\sltwo$ are pointlike at the center $\rho=\tau=0$ of $AdS_3$.  They are the analogue of ZZ branes in Liouville theory, which when applied to the 2d noncritical string%
~\rcite{McGreevy:2003kb,Douglas:2003up,Balthazar:2019rnh} 
describe dynamics in the strong coupling regime.%
\footnote{The $AdS_2$ branes we have been discussing are the analogue of FZZT branes in the 2d noncritical string (see~\rcite{Balthazar:2018qdv} for a recent discussion and further references).}
In the present context, the $\sltwo$ identity brane will arise when incoming scattering states access the strong-coupling region of the winding tachyon condensate in the center of $AdS_3$~\rcite{Martinec:2021vpk}.  Rather than turning the spacetime CFT into a BCFT, such ``identity branes'' describe nonperturbative dynamical processes even in absence of a defect brane.

Another class of $AdS_3$ D-brane is the instantonic $dS_2$ brane~\rcite{Bachas:2000fr,Quella:2002ns,Israel:2005fn,Israel:2005ek}.  The recent analysis of these branes in~\rcite{Gaberdiel:2021kkp} is somewhat complementary to the present work; their focus was on the symmetric orbifold for $k=1$ in the critical dimension, nevertheless one might expect that much of their analysis of $dS_2$ branes should carry over to the noncritical $k<1$ models. 

In addition to these symmetry-preserving D-branes, there is a class of symmetry-breaking branes that preserve only a diagonal $\uone$ instead of a diagonal $\sltwo$ (or $\sutwo$) on the worldsheet boundary~\rcite{Maldacena:2001ky,Quella:2002ct,Walton:2002db,Sarkissian:2002ie,Sarkissian:2002bg,Quella:2002ns,Quella:2003kd}; their application to $AdS_3$ string theory is discussed in~\rcite{Martinec:2019wzw}.  The reduction of the symmetry group has the effect of smearing the symmetry-preserving branes along an orbit of the group.  For instance, the symmetry-breaking brane built on the $AdS_2$ brane smears it azimuthally so that fills $AdS_3$ outside the radius $\rho_{\rm min}=\mu$.  Open strings can then end anywhere on this smeared conjugacy class.  There are now no ``sides'' of the brane, since it fills the $AdS_3$ boundary; but there are long open strings whose endpoints lie on this brane.  In this case, one will have not a defect CFT in spacetime but rather some sort of BCFT with a boundary degree of freedom specifying where azimuthally on $AdS_3$ a given long open string endpoint asymptotes to on the conformal boundary.

Starting instead from the perspective of the spacetime CFT, closely related to the present work are recent studies of boundary states in symmetric products~\rcite{Belin:2021nck}.  As mentioned in section~\ref{sec:OpenSymProd}, general open string boundary conditions do not correspond to a bulk theory of perturbative strings, in accord with the conclusions of~\rcite{Belin:2021nck}. 

Another aspect we have barely touched upon in this work is the interplay between Euclidean and Lorentzian $AdS_3$.  One often thinks of the Euclidean theory as preparing an initial state in the Lorentzian continuation; indeed, as discussed in~\rcite{Balthazar:2021xeh} the perturbative worldsheet amplitudes might be most naturally thought of as calculating Euclidean correlators.  D-branes in the Euclidean theory might then be naturally thought of as preparing coherent highly excited states in the theory; this might be a useful way of exploring this regime using perturbative methods.  We hope to return to this issue in future work.

The deformed symmetric product structure is well-suited for describing the long string continuum of fundamental strings in $AdS_3$.  For $k<1$, it is believed that this sector constitutes the leading contribution to the entropy, and so the deformed symmetric product is a candidate for the complete spacetime CFT dual.  For $k>1$, a proposal has been put forth~\rcite{Eberhardt:2021vsx} for a symmetric product describing the long string sector, with an impressive matching of conformal perturbation theory in the interaction deformation up to third order between the worldsheet theory and the candidate spacetime theory.  However, this symmetric product cannot be the whole story, since for $k>1$ there is a BTZ black hole spectrum not captured by the Fock space of perturbative long strings (and relatedly, the full spacetime CFT has an $\sltwo$ invariant vacuum, which the candidate spacetime CFT does not).  It was argued in~\rcite{Balthazar:2021xeh} that this model only captures the long string dynamics in some finite range of radial position where the long strings are weakly coupled.  Nevertheless, the results of~\rcite{Eberhardt:2021vsx} suggest that there might be a scaling limit where one takes $p\to\infty$ while also scaling the strength $\mu$ of the twist interaction to zero, holding some relation fixed; this $c_\ST\to\infty$ limit would push the BTZ threshold off to infinite energy, while decoupling the light excitations near the vacuum at the center of $AdS_3$, and might thus permit a consistent theory.  If such a limit exists, it should straightforwardly extend to the BCFT construction above.


\vskip .8cm
\noindent
\ref{sec:discussion}.5~~{\em Similarities to $c=1$ noncritical strings}
\medskip

In the 2d noncritical string (see \eg~\rcite{Martinec:2004td} for a review), the worldsheet description involves a target space $\bR_\phi\times \bR_t$, with a linear dilaton in the spatial direction $\bR_\phi$ and an exponential (Liouville) potential keeping strings out of strong coupling at low energies.  The dual theory is a matrix quantum mechanics in an inverted oscillator potential, in which the light excitations are perturbations of the matrix eigenvalue distribution.  The individual eigenvalues are D0-branes in the worldsheet description, which manifest themselves in the strong coupling region both as the leading contribution to certain non-perturbative processes (matrix eigenvalues tunneling through the inverted oscillator barrier), and as unstable objects (matrix eigenvalues well-separated from the eigenvalue condensate) that decay into closed string radiation.

The structures of $AdS_3/CFT_2$ at $k<1$ and 2d noncritical string theory are quite parallel.  D-branes that are the tensor product of identity branes in $\sltwo$ times an $\bS^2_\flat$ brane in $\bS^3_\flat$ describe Euclidean tunneling processes in the worldsheet expansion of the long string S-matrix; and the symmetry-breaking version of the identity brane is an unstable Lorentzian D-brane that will decay into closed string radiation.
These branes were described in~\rcite{Martinec:2019wzw,Martinec:2021vpk}, where they were related to little string theory excitations of the underlying background NS5-branes.
The extent of this similarity leads one to wonder whether there is a yet further layer underlying $AdS_3/CFT_2$ duality in the regime $k<1$, in which the long strings are themselves collective excitations of a little string fluid associated to a condensate of ``identity branes'' in $AdS_3$, just as closed strings in the 2d noncritical string are collective excitations of a D0-brane fluid.

While the theories defined here and in~\rcite{Balthazar:2021xeh} are consistent at finite fundamental string charge $p$, they generically describe scattering states of long strings, and as such a packet of strings that enter the fivebrane throat, reach some maximum depth, and then scatter back out.  The situation here is again somewhat similar to the $c=1$ matrix model, which at finite $N$ describes a blob of D0-branes that scatter off the matrix potential and back out, leaving nothing behind~\rcite{Karczmarek:2003pv}.  A steady state for the $k<1$ theory may generically require feeding in long strings at the same rate they are returned to the asymptotic region, just as in the $c=1$ matrix model one must keep feeding in D0-branes in order to maintain a static background.



\vskip 1cm

\section*{Acknowledgements}

Thanks go to 
Bruno Balthazar, 
David Kutasov,
and
Edward Mazenc
for useful discussions.
This work is supported in part by DOE grant DE-SC0009924.


\appendix


\section{Conventions}
\label{sec:conventions}


\subsection{\texorpdfstring{$SU(2)$}{}}
\label{sec:sutwo}

The $N=1$ superconformal $\sutwo_n$ WZW model consists of a bosonic $\sutwo_{n-2}$ WZW model, with currents $j^a_\su(z)$, and three free fermions $\psi^a_\su$, $a=3,+,-$ from which one can construct a $SU(2)_2$ current algebra. The total central charge is given by
\eqn[csu2]{c_{SU(2)}=c^{\rm bos}+c^{\rm ferm}=\frac{3(n-2)}{n}+\frac{3}{2}~.}
The bosonic currents and free fermions satisfy the OPE's
\eqn[bbhh]{
j^a_\su (z)\, j^b_\su(0) \sim \frac{n-2}{2 z^2}\,\delta^{ab}+i\epsilon^{ab}_{~\;c}\,\frac{j^c_\su(0)}{z}
~~,~~~~
\psi^a_\su(z)\psi^b_\su(0)\sim\frac{\delta^{ab}}{z}~.
}
The total $\sutwo$ currents are 
\eqn[bbii]{
J^a_\su=j^a_\su-\frac{i}{2}\epsilon^{a}_{~bc}\psi^b_\su\psi^c_\su ~.
}
This theory has $N=1$ superconformal symmetry with supercurrent
\eqna[Gsu]{G_\su&=\sqrt{\frac{2}{n}}\left(\psi^a_\su \, j^a_\su -i\psi^1_\su\psi^2_\su\psi^3_\su\right) \cr &=\sqrt{\frac{1}{n}}\left(\psi^+_\su j^-_\su+\psi^-_\su j^+_\su\right)+\sqrt{\frac{2}{n}}J^3_\su\psi^3_\su ~,}
where 
\eqna[jpsipm]{j^\pm_\su=j^1_\su\pm i j^2_\su,~~~~\psi^\pm_\su=\frac{\psi^1_\su\pm i\psi^2_\su}{\sqrt{2}}~.}
The fermions $\psi^a_\su$ and total currents $J^a_\su$ form $N=1$ supermultiplets of dimension $(\half,0)$ under~\eqref{Gsu}.

The current algebra primaries of the bosonic $\sutwo_{n-2}$ are given by the operators $v_{j';m',\mbar'}$, which have conformal weight
\eqn[hvjm]{h\left[v_{j';m',\mbar'}\right]=\frac{j'(j'+1)}{n} ~,}
and satisfy the following OPE's with the bosonic $SU(2)$ currents (in a particular normalization of the $v$'s),
\eqna[su2reps]{&j^3_\su(z) v_{j';m',\mbar'}(0)\sim \frac{m}{z} \, v_{j';m',\mbar'}(0)~,\cr
&j^\pm_\su (z)\, v_{j';m',\mbar'}(0)\sim \frac{\sqrt{j'(j'+1)-m'(m'\pm 1)}}{z} \, v_{j';m'\pm1}(0)~.}
Unitary representations lie in the range $j'=0,\half, 1,\cdots, \frac n2-1$; $m',\bar m'=-j', -j'\tight+1,\cdots, j'$.

The bosonic $\sutwo_{n-2}$ WZW theory admits a spectral flow transformation under which the modes of the currents transform as\footnote{In some of the equations below we will drop the subscript ``$\su$'' to avoid clutter.}
\eqna[sutwospecflow]{
j^\pm_n \to j^\pm_{n\pm w'}
~&,~~~~
j^3_n\to j^3_n + \frac{\ntil}2w' \, \delta_{n,0}~,
\\
\bar j^\pm_n \to \bar j^\pm_{n\pm \bar w'}
~&,~~~~
\bar j^3_n\to \bar j^3_n + \frac{\ntil}2\bar w' \, \delta_{n,0}~,
}
with $w',\bar w'\in\bZ$ obeying the restriction $w'-\bar w'\in 2\bZ$. Under spectral flow,
the operator $v_{j';m',\mbar'}$ flows to an operator 
$v^{(w',\bar w')}_{j';m',\mbar'}$ 
which has conformal weight and $j^3$ charge
\eqn[sutwoqnos]{h\big[v^{(w',\bar w')}_{j';m',\mbar'}\big]= \frac{j'(j'+1)}{n} + m' w' + \frac{\ntil}{4}(w')^2
~~,~~~~
j_0^3\big[v^{(w',\bar w')}_{j';m',\mbar'}\big]=m'+\frac{\ntil}{2}w'~, }
and similarly for the right-moving spectral flow.  The flow is an automorphism of the affine Lie algebra, that takes the highest weight state to current algebra descendants.  One can understand it via the decomposition of $v_{j';m',\mbar'}^{(w',\bar w')}$ into its $\uone$ and $\frac\sutwo\uone$ components 
\eqna[bospf]{
v_{j';m',\mbar'}^{(w',\bar w')} = \lambda_{j';m',\mbar'} \,
\exp\left[i\sqrt{\frac{2}{n-2}}\left(\Big(m'+\frac{n-2}{2}\,w'\Big)\,y_\su + 
\Big(\mbar'+\frac{n-2}{2}\,\bar w'\Big)\,\bar y_\su\right) \right]~,
}
where $y_\su,\bar y_\su$ bosonize the $\uone$ currents
\eqn[jbos]{
j^3_\su=i\sqrt{\frac{n-2}{2}}\, \partial y_\su
~~,~~~~
\bar j^3_\su=i\sqrt{\frac{n-2}{2}}\, \bar\partial \bar y_\su ~,
}
and $\lambda_{j';m',\mbar'}$ is an operator in the $\frac\sutwo\uone$ coset model (and thus neutral under the $\uone$ currents $j^3_\su,\bar j^3_\su$).  This {\it parafermion decomposition}
shows that the spectral flow quanta $w',\bar w'$ act as ``winding numbers'' conjugate to the left and right zero mode momenta $m',\mbar'$ (though of course there is no conserved winding number in $\sutwo=\bS^3$).
One can see from \eqref{bospf} that the operator $v^{(w',\bar w')}_{j';m',\mbar'}$ is a Virasoro primary for any $w',\bar w'\in \bZ$, but in general it is not a current algebra primary.

In the supersymmetric theory, one can also perform spectral flow with respect to the total $SU(2)$ algebra, $J^a_n$. This combines the bosonic spectral flow described above with a spectral flow for the fermions.  To describe it explicitly, it is convenient to bosonize the fermions $\psi^\pm_\su$: 
\eqn[sutwobos]{
\psi^+_\su\psi^-_\su = i\partial H_\su 
~~,~~~~
\psi^\pm_\su = e^{\pm i H_\su} ~,
}
and similarly for $\bar\psi^\pm_\su$. Here $H_\su$ is normalized in the standard CFT way, $H_\su(z)H_\su(w)\sim -\ln(z-w)$. We make the same choice for other scalar fields that appear in our construction, such as $y_\su$ \jbos.

The supersymmetric spectral flow takes the operator $v_{j';m',\mbar'}$ to
\eqn[aarr]{V^{(w',\bar w')}_{j';m',\mbar'}=e^{i w' H_\su+i\bar w'\bar H_\su} \; v^{(w',\bar w')}_{j';m',\mbar'}~.} 
Using the superconformal current of $\sutwo_n$ given by~\eqref{Gsu}, one can check that $V^{(w',\bar w')}_{j';m',\bar{m}'}$ is a superconformal primary (but, again, not a current algebra primary). 

The supersymmetric $SU(2)_n$ WZW model admits a super-parafermionic decomposition in terms of an $N=2$ supersymmetric $\frac{SU(2)_n}{U(1)}$ coset CFT and a free superfield $(\psi^3_\su, J^3_\su)$. The spectral flow \eqref{aarr} does not act on the $\frac{SU(2)_n}{U(1)}$ coset, but only on the $U(1)$ part. 

The various $U(1)$ currents can be bosonized as~\eqref{jbos}, \eqref{sutwobos}, as well as
\eqna[su2bos]{
&J^3_\su=i\sqrt{\frac{n}{2}}\,\partial Y
~~,~~~~
J^R_\su=i\sqrt{\frac{n-2}{n}}\,\partial Z \equiv ia\,\partial Z~,}
where $J^R_\su$ is the $U(1)$ R-symmetry current of the $N=2$ supersymmetric $\frac{SU(2)_n}{U(1)}$ coset CFT,
\eqn[JRsu2]{J^R_\su=\psi^+_\su\psi^-_\su+\frac{2}{n}J^3_\su ~.}
Note that it is orthogonal to $J^3_\su$, as is implied by the decomposition of the SCFT $SU(2)_n$ described above.

The scalar fields defined in~\eqref{jbos}, \eqref{sutwobos}, \eqref{su2bos} are related by a field space rotation
\eqna[YHrot]{
Y &= \sqrt{\frac{n-2}{n}} \, y_\su + \sqrt{\frac2n}\, H_\su~,
\\
Z &= -\sqrt{\frac2n}\,y_\su + \sqrt{\frac{n-2}{n}}\, H_\su ~.
}
In this rotated basis, the $SU(2)_n$ operator with general fermion charges $\eta_\su,\bar\eta_\su$ and bosonic spectral flow $w',\bar w'$
\eqna[Vparaf]{V^{(\eta_\su,\bar\eta_\su,w',\bar{w'})}_{j';m',\bar{m}'}&\equiv e^{i\eta_\su H_\su+i\bar{\eta}_\su\bar{H}_\su}v^{(w',\bar{w'})}_{j';m',\bar{m}'}}
can be decomposed in terms of an exponential operator carrying the total $J^3_\su$ charge, together with a charge-neutral super-parafermion operator
\eqna[pfdecomp]{
V_{j';m',\mbar'}^{(\eta_\su,\bar\eta_\su,w',\bar w')} &=  
 \Lambda_{j';m',\mbar'}^{(\alpha,{\bar\alpha})} \,\exp\Bigl[ i\sqrt{\frac{2}{n}} \Bigl(m'+\alpha+\frac{n}{2} w'\Bigr)Y
+ i\sqrt{\frac{2}{n}} \Bigl(\mbar'+{\bar\alpha}+\frac{n}{2} \bar w'\Bigr)\bar Y \Bigr]~, }
where $\etatil=\eta_\su-w'$, $\bar\alpha=\bar\eta_\su-\bar w'$, and $\Lambda_{j;m',\mbar'}^{(\alpha,{\bar\alpha})} $ is the super-parafermion operator of the supersymmetric $\frac{SU(2)_n}{U(1)}$ coset CFT, whose left and right scaling dimensions are
\eqna[pfspec]{
h\big[\Lambda_{j';m',\mbar'}^{(\alpha,\bar\alpha)}\big] &= \frac{j'(j'+1)}{n}-\frac{(m'+\alpha)^2}{n}+\frac{ \alpha^2}{2} ~,
\\
\bar h\big[\Lambda_{j';m',\mbar'}^{(\alpha,\bar\alpha)}\big]  &= \frac{j'(j'+1)}{n}-\frac{(\mbar'+\bar{\alpha})^2}{n}+\frac{ \bar{\alpha}^2}{2}  ~.
}
The super-parafermion operators can be further decomposed in terms of an exponential carrying the R-charge and the bosonic parafermion introduced in~\eqref{bospf}
\eqna[bospf2]{
\Lambda_{j';m',\mbar'}^{(\alpha,{\bar\alpha})} &= 
\lambda_{j';m',\mbar'}^{~} \;\exp\left[ i\frac{2}{\sqrt{n(\ntil)}} \left(
\Bigr(-m'+\frac{\ntil}{2}\alpha\Bigr) Z 
+  \,\Bigr(-\mbar'+\frac{\ntil}{2}{\bar\alpha}\Bigr) \bar Z\right)\right] ~.
}
We will use interchangeably the descriptors supersymmetric $\frac\sutwo\uone$ coset model, $N=2$ minimal model, $N=2$ Landau-Ginsburg model, and parafermion theory to refer to the same CFT.

Another essential aspect of the $SU(2)_n$ theory is the moduli space of deformations by the operator $\int\! d^2z J^3_{su} \bar{J}^3_{su}$.  This deformation clearly does not act on the $\frac\sutwo\uone$ part of the operators, \eg\ on the operators $\Lambda_{j';m',\mbar'}^{(\alpha,\bar\alpha)}$ in \eqref{pfdecomp}. The action of the deformation on the $U(1)$ factor corresponds to changing the radius of $Y$.  The parafermion decomposition~\eqref{pfdecomp} is then modified to
\eqn[pfdecompsq]{
V_{j';m',\mbar'}^{(\eta_\su,\bar\eta_\su,w',\bar w')} =  
\Lambda_{j';m',\mbar'}^{(\alpha,{\bar\alpha})} \,
\exp\Bigl[ i \Bigl(p_Y Y + \bar p_Y\bar Y\Bigr) \Bigr]~. }
The spectrum of the deformed exponential in $Y,\bar Y$ can be written as~\rcite{Yang:1988bi,Giveon:1993ph,Brennan:2020bju} 
\eqna[squonespec]{
(h,\bar h) &= \half(p_Y^2,\bar p_Y^2)~,
\\
\left(p_{Y}^{~},\bar p_Y^{~}\right) &= \frac{1}{\sqrt{2n}}\Big(\frac{p+nP}{R}\pm R(\ell+nL)\Big)~,
}
with the quantum numbers related to those of~\eqref{pfdecomp} via
\eqna[paramrel]{
2(m'+\alpha) = p+\ell ~~&,~~~~  w' = P+L
\quad,\qquad P,L\in\bZ~,
\\
2(\bar m'+{\bar\alpha}) = p-\ell ~~&,~~~~\bar w' = P-L 
\quad,\qquad p,\ell \in\{0,...,n\tight-1\} ~.
}
Here, $\alpha,{\bar\alpha}$ are spectral flow parameters under the $N=2$ superconformal symmetry of the $\frac\sutwo\uone$ coset model, and $w',\bar w'$ are spectral flow parameters in the $\uone$ CFT.  Essentially, the circular orbits of the vector $\uone$ isometry of $\sutwo$ (generated by $J^3_\su+\bar J^3_\su$) expand in size by a factor $R$, while the circular orbits of the corresponding axial $\uone$ isometry (generated by $J^3_\su-\bar J^3_\su$) shrink by the same factor (for a discussion in the present context, see~\rcite{Brennan:2020bju}).  We refer to this deformed geometry as a squashed $\bS^3$, and denote it by $\sqsphere$.

\vskip 1cm

\subsection{\texorpdfstring{$SL(2,\bR)$}{}}
\label{sec:sltwo}

The above discussion can be repeated, with a few interesting twists, for the case of $\sltwo$. The supersymmetric $\sltwo_k$ WZW model consists of a bosonic $\sltwo_{k+2}$ WZW model, with bosonic currents $j^a_\sl(z)$, and three free fermions $\psi_\sl^a$, which give a $\sltwo_{-2}$ affine Lie algebra. The total central charge is given by
\eqn[csl2]{c_{SL(2)}=c^{\rm bos}+c^{\rm ferm}=\frac{3(k+2)}{k}+\frac{3}{2} ~.}

The bosonic currents and free fermions satisfy the OPE's
\eqn[aazz]{j^a_\sl(z)j^b_\sl(0)\sim\frac{k+2}{2 z^2}\eta^{ab}+i\epsilon^{ab}_{~\;c}\frac{j^c_\sl(0)}{z}~,~~~\psi^a_\sl(z)\psi^b_\sl(0)\sim\frac{\eta^{ab}}{z} ~,}
where $\eta^{ab}=(+,+,-)$, and $\epsilon^{123}=1$. The total currents
\eqn[bbaa]{J^a_\sl=j^a_\sl-\frac{i}{2}\epsilon^a_{~bc}\psi^{b}_\sl\psi^c_\sl}
have level $(k+2)+(-2)=k$.
The theory has $N=1$ superconformal symmetry with supercurrent
\eqna[Gsl]{G_\sl&=\sqrt{\frac{2}{k}}(\eta_{ab}\,\psi^a_\sl \,j^b_\sl+i\psi^1_\sl\psi^2_\sl\psi^3_\sl)\cr &=\sqrt{\frac{1}{k}}\left(\psi_\sl^+ j^-_\sl+\psi_\sl^- j^+_\sl\right)-\sqrt{\frac{2}{k}}J^3_\sl\psi_\sl^3 ~,}
where 
\eqn[jpsisl]{j^\pm_\sl=j^1_\sl\pm i j^2_\sl,~~~\psi^\pm_\sl=\frac{\psi^1_\sl\pm i \psi^2_\sl}{\sqrt{2}}~.}
Primary operators under the bosonic $\sltwo_{k+2}$ current algebra are operators $\Phi_{j;m,\mbar}^{(0)}$, which have conformal weight
\eqn[bbbbgggg]{h\big[\Phi_{j;m,\mbar}^{(0)}\big]=-\frac{j(j-1)}{k} ~,}
where the meaning of the superscript $(0)$ will be clarified below.  These operators satisfy the OPE's with respect to the bosonic $\sltwo$ currents
\eqna[sl2reps]{&j^3_\sl(z) \Phi_{j;m,\mbar}^{(0)}(0)\sim \frac{m}{z}\Phi_{j;m,\mbar}^{(0)}(0)~,\cr
&j^\pm_\sl(z) \Phi_{j;m,\mbar}^{(0)}(0)\sim \frac{m\mp (j-1)}{z}\Phi_{j;m\pm1,\mbar}^{(0)}(0) ~.}
Some of the operators $\Phi^{(0)}_{j;m,\mbar}$ correspond to normalizable or delta-function normalizable states.  Normalizable states belong to unitary discrete representations $\cD_j^\pm$ of the bosonic theory, known as the principal discrete series, for which $j$ lies in the range 
\eqn[unitaryrangeSL]{
\half<j<\half\big( k+1 \big) ~,
}
with $m-j\in\bN_0$ for $\cD^+$, and $-j-m\in\bN_0$ for $\cD^-$ (where $\bN_0$ are the non-negative integers).  Delta-function normalizable states belong to the principal continuous series $\cC_{j,\alpha}$, $j\in \half+i\bR$ and $\alpha\in(0,1)$.  In string theory, the principal discrete series representations $\cD_j^-$ describe in-states for normalizable states of strings in $AdS_3$, while $\cD_j^+$ describes the corresponding out-states.  One can think of the two types of representations as bound states and scattering states, respectively.  Note that the $AdS_3$ vacuum corresponds to $j=1$ and is thus non-normalizable for $k<1$, according to \unitaryrangeSL.

In CFT and string theory on $AdS_3$ one also needs to consider non-normalizable operators, that do not satisfy the bound \eqref{unitaryrangeSL}. These operators give rise to local operators in the spacetime CFT~\rcite{Kutasov:1999xu}.  A convenient semi-classical description of such operators in Euclidean $AdS_3=\frac{\sltwoc}{\sutwo}\equiv \bH_3^+$ parametrizes the target space via the matrix~\rcite{Teschner:1997fv}
\eqn[hmat]{
h = 
\left(\begin{matrix} 1 & 0 \\ \gamma & 1 \end{matrix}\right)
\left(\begin{matrix}
e^\phi & 0 \\
0 & e^{-\phi}  
\end{matrix}\right)
\left(\begin{matrix} 1 & \bar\gamma \\ 0 & 1 \end{matrix}\right)
=
\left(\begin{matrix}
e^\phi & e^\phi \bar\gamma \\
e^\phi \gamma & e^{-\phi} + e^{\phi} \gamma\bar\gamma
\end{matrix}\right)~,
}
on which $g\in\sltwoc$ acts via $h\to g^{-1}h(g^{-1})^\dagger$.  The functions
\be\label{scalingfns}{
\Phi_j(x,\bar x) = \frac{2j-1}{\pi}\biggl(\bigl(x,1)\cdot h\cdot \biggl(\begin{matrix} \bar x \\ 1\end{matrix}\biggr)\biggl)^{-2j}
= \frac{2j-1}{\pi} \Big( |\gamma-x|^2e^{\phi} + e^{-\phi} \Big)^{-2j}
}\ee
are eigenfunctions of the Laplacian on $\bH_3^+$. The complex parameter $x$ labels points on the boundary. $\Phi_j(x,\bar x)$ transforms as a tensor of weight $(j,j)$ under $\sltwoc$.
In CFT on $AdS_3$, the functions \eqref{scalingfns} are promoted to operators $\Phi_j(x,\bar x;z,\bar z)$, as their arguments $\phi,\gamma,\bar\gamma$ are now two dimensional fields. 

The $\sltwo$ currents can also be written in the position basis on the boundary, as 
\eqna[Jofx]{
j(x;z) &= -j^+_\sl(z)+2x j^3_\sl(z)-x^2 j^-_\sl(z) ~, 
\\[.1cm]
\psi(x;z) &= -\psi^+_\sl(z)+2x \psi^3_\sl(z)-x^2 \psi^-_\sl(z)  ~,
\\
J(x;z) &= j(x;z) + \half\psi(x;z)\partial_x\psi(x;z) ~.
}
The current algebra and the transformation properties of the local operators $\Phi_j(x,\bar x;z,\bar z)$ are given in this basis by 
\eqna[JPhialgebra]{
J(x;z)J(y;w) &\sim k\frac{(y-x)^2}{(z-w)^2} + \frac{1}{z-w}\big[(y-x)^2\partial_y-2(y-x)\big] J(y;w)~,
\\
J(x;z)\Phi_h(y,\bar y;w,\bar w) &\sim \frac{1}{z-w}\big[(y-x)^2\partial_y + 2h(y-x)\big]\Phi_h(y,\bar y; w,\bar w)~.
}
At large $\phi$ (\ie\ near the boundary of $AdS_3$), the operators $\Phi_j$ behave as 
\eqn[Phiasymp]{
\Phi_j(x,\bar x) \sim
e^{(j-1)Q\phi}\, \delta^2(\gamma-x)+\cO\big(e^{(j-2)Q\phi)}\big) 
+ \frac{2j-1}{\pi} \frac{e^{-jQ\phi}}{|\gamma-x|^{4j}} + \cO\big(e^{-(j+1)Q\phi}\big) 
+\cdots~,
}
where $Q=\sqrt{2/k}$, $\phi$ has been rescaled by $Q/2$ relative to~\eqref{hmat}, \eqref{scalingfns}, and the meaning of the ellipses will be explained below. For $j>1/2$, the operator~\eqref{Phiasymp} is non-normalizable due to a divergence of the corresponding wavefunction as $\phi\to\infty$, in which case the leading term in~\eqref{Phiasymp} shows that $\Phi_j$ reduces to a local operator on the conformal boundary.%
\footnote{
The norm, given by 
$$
\int\! d\phi d\gamma d\bar\gamma \,e^{Q\phi}\, \big|\Phi_j \big|^2 \sim \int\! d\phi \, e^{Q(2j-1)\phi} \,(\cdots) ~,
$$
shows that the operators $\Phi_j$ in~\eqref{Phiasymp} are non-normalizable for $j>\half$.
}

The operators $\Phi_j$ obey a reflection symmetry~\rcite{Teschner:1997fv,Giveon:2001up}
\eqn[reflection]{
\Phi_j(x,\bar x) = \frac{2j-1}{\pi} \int\! d^2 x' \, |x-x'|^{-4j}\, \Phi_{1-j}(x',\bar x')~,
}
which for real $j$ allows us to restrict our attention to $j>\hf$, and for $j=\hf+is$ says that the operators with $s>0$ and $s<0$ are not independent, as one would expect. 

For some purposes it is convenient to transform the local operators $\Phi_j(x,\bar x;z,\bar z)$ from position space $(x,\bar x)$ to momentum space $(m,\bar m)$, as in~\eqref{bbbbgggg}, \eqref{sl2reps}.  These representations are related by
\eqna[Phiexpn]{
\Phi_{j;m,\mbar}^{(0)} &= \int\! d^2x\, x^{j+m-1}\bar x^{j+\mbar-1} \, \Phi_{j}(x,\bar x) ~.
}
Plugging the asymptotic expansion \eqref{Phiasymp} into \eqref{Phiexpn}, we find that the operators $\Phi_{j;m,\mbar}^{(0)}$ behave at large $\phi$ like
\eqna[Phimodeasymp]{
\Phi_{j;m,\mbar}^{(0)} &\sim
e^{(j-1)Q\phi} \,\gamma^{j+m-1}\bar\gamma^{j+\mbar-1}+\cO(e^{(j-2)Q\phi})  \\
&\hskip .5cm
+(2j\tight-1) \frac{\Gamma(j\tight+m)\Gamma(j\tight-\mbar)\Gamma(1\tight-2j)}{\Gamma(m\tight-j\tight+1)\Gamma(-\mbar\tight-j\tight+1)\Gamma(2j)}
\,e^{-jQ\phi} \, \gamma^{m-j}\bar\gamma^{\mbar-j} + \cO(e^{-(j+1)Q\phi}) +\cdots ~.
}
The momentum variables $(m,\bar m)$ must satisfy the constraint $m-\mbar\in\bZ$, necessary for the single-valuedness of the integral \eqref{Phiexpn} (but are otherwise unrestricted and in particular unrelated to $j$); $m-\bar m$ is the spatial momentum on the boundary, which is quantized since the spatial coordinate lives on a circle.

From the expansion \eqref{Phimodeasymp} we see that for general $j$, when the momentum variables $(m,\bar m)$ take some specific values, the coefficient of the leading decaying (normalizable) term diverges. An example is $-(m+j)\in \bN_0$.
These divergences correspond to values of $j,m$ at which the local, non-normalizable operator $\Phi_j$ can create a normalizable state from the vacuum. It is an analog of the LSZ reduction in standard QFT, and is discussed in detail in a closely related context (the $\frac{\sltwo}{U(1)}$ coset) in~\rcite{Aharony:2004xn}. That paper also discusses other singularities of \eqref{Phimodeasymp} that do not have this interpretation.

To implement the above procedure in our case we proceed as follows. Consider the case $m=\bar m$, $m+j\to 0$, as an example. The operator \eqref{Phimodeasymp} diverges in this limit; therefore, we need to take the limit more carefully, to ensure that the operator remains finite. To do that, we define
\be
\label{modphiop}
\widetilde{\Phi}_{j;-j,-j}^{(0)}\equiv\lim_{m,\bar m\to -j}(m+j)\Phi_{j;m,\mbar}^{(0)}.
\ee
Looking back at the expansion \eqref{Phimodeasymp}, we see that in this limit the leading, non-normalizable, contribution to $\Phi_{j;m,\mbar}^{(0)}$ disappears, and we are left with a finite operator that behaves at large~$\phi$ like $e^{-Qj\phi}$. For $j>\half$ this operator is normalizable, and thus it describes a normalizable state, at least semi-classically. 

In the exact theory the situation is more interesting. The FZZ correspondence~\rcite{FZZref,Giveon:1999px,Maldacena:2000hw,Kazakov:2000pm,Giveon:2016dxe,Martinec:2020gkv} asserts that the normalizable operator \eqref{modphiop} has additional contributions from sectors with non-zero winding associated to the ellipses in equations~\eqref{Phiasymp}, \eqref{Phimodeasymp}. The question whether the operator \eqref{modphiop} is indeed normalizable or not depends on the nature of these contributions. In the semi-classical limit $k\to\infty$ with $j$ fixed, they are known to be rapidly decaying at large $\phi$, and thus the operator \eqref{modphiop} is normalizable. If $k$ is large but $j$ scales like $k$, they actually can be dominant at large $\phi$, and can even make this operator non-normalizable. The condition that the operator remains normalizable is in fact the origin of the upper bound on $j$ in equation \eqref{unitaryrangeSL}, which is valid for arbitrary $k$.

Another correction to the semi-classical picture above in the full quantum theory, which is related to the one we just discussed is the fact that the reflection symmetry \eqref{reflection} is modified~-- the RHS of \eqref{reflection} is multiplied by a factor $\cR(j)$, with
\eqn[Rj]{\cR(j)=\frac{\Gamma\left(1-\frac{2j-1}{k}\right)}{\Gamma\left(1+\frac{2j-1}{k}\right)} ~.}
This factor goes to one in the classical limit $k\to\infty$, and for the delta-function normalizable case is a pure phase. This modification means that the subleading terms in \eqref{Phiasymp}, \eqref{Phimodeasymp} are multiplied by $\cR(j)$ as well.  


The bosonic $\sltwo$ CFT again admits a spectral flow transformation under which the currents transform as
\eqna[sltwospecflow]{
j^\pm_n \to j^\pm_{n\pm w}
~&,~~~~
j^3_n\to j^3_n + \frac{\ktil}2w \, \delta_{n,0}~,
\\
\bar j^\pm_n \to \bar j^\pm_{n\pm w}
~&,~~~~
\bar j^3_n\to \bar j^3_n + \frac{\ktil}2 w \, \delta_{n,0}~,
}
where $w$ is an integer, and we have dropped the subscript ``$\sl$'' to avoid clutter. The operator $\Phi_{j;m,\mbar}^{(0)}$ flows to an operator $\Phi^{(w)}_{j;m,\mbar}$ which has conformal weight and $j^3$ charge
\eqn[aaqq]{h[\Phi^{(w)}_{j;m,\mbar}]=-\frac{j(j-1)}{k}-m w-\frac{\ktil}{4}w^2,~~~j_0^3[\Phi^{(w)}_{j;m,\mbar}]=m+\frac{\ktil}{2}w ~.}
The operator $\Phi^{(w)}_{j;m,\mbar}$ is again a Virasoro primary for any $w\in \bZ$, but it is not a current algebra primary. The OPE's~\eqref{sl2reps} generalize to
\eqna[sl2windingreps]{&j^3_\sl(z) \Phi^{(w)}_{j;m,\mbar}(0)\sim \frac{m+\frac{k+2}{2}w}{z}\Phi^{(w)}_{j;m,\mbar}(0)~,\cr
&j^\pm_\sl(z) \Phi^{(w)}_{j;m,\mbar}(0)\sim \frac{m\mp (j-1)}{z^{\pm w+1}}\Phi^{(w)}_{j;m\pm1,\mbar}(0) ~.}
Note that the left and right spectral flows in \eqref{sltwospecflow} are identical because we are working on the universal cover of $\sltwo$. Thus, the timelike direction in the group manifold is non-compact and has no winding.  In contrast to the $\sutwo$ case, this spectral flow is not an automorphism of representations, rather it generates new representations of the current algebra~\rcite{Maldacena:2000hw}.

As mentioned above, string winding number is not conserved, and so operators described in a given sector have contributions from other sectors.  Related to this is the phenomenon of FZZ duality~\rcite{FZZref,Giveon:1999px,Maldacena:2000hw,Kazakov:2000pm,Giveon:2016dxe,Martinec:2020gkv} which relates $\cD^-$ unitary representations in winding sector $w$ to $\cD^+$ unitary representations in winding sector $w-1$; in particular highest weight states are identified via
\eqn[FZZduality]{
\Phi_{j;-j,-j}^{(w)} \equiv \Phi_{\frac k2+1-j;\frac k2+1-j,\frac k2+1-j}^{(w-1)} ~.
}
Note from equation~\eqref{aaqq} that the conformal dimension and $j^3$ eigenvalues match. The rest of the map between the representations follows from the spectral flow of the generators~\eqref{sltwospecflow}.  
Thus for instance when we are counting states we should not include both sets of representations $\{\cD^+_{(w),j}\}$ and $\{\cD^-_{(w),j}\}$, but only one or the other.  Our conventions are such that $\cD^-_{(w),j}$ representations with $w\le -1$ describe in-states bound to $AdS_3$ that wind $|w|$ times around the azimuthal direction.  The remaining $\cD^-_{(w),j}$ representations with $w\ge0$ map via FZZ duality to the set of $\cD^+_{(w),j}$ representations with $w\ge1$, which are charge conjugates of the $\cD^-_{(w),j}$ representations with $w\le-1$, and which thus describe out-states.

As in the $SU(2)$ case, in the supersymmetric case we can also consider spectral flow with respect to the total $\sltwo_{k}$ algebra. To do this, we combine the operators $\Phi^{(w)}_{j;m,\mbar}$ with a contribution from the fermions
\eqn[sltwobos]
{
\psi^+_\sl\psi^-_\sl = i\partial H_\sl 
~~,~~~~
\psi^\pm_\sl = e^{\pm i H_\sl} ~,
}
and consider the operators 
\eqn[superPhi]{
\Phihat^{(w)}_{j;m,\mbar} = e^{-i w (H_\sl + \bar H_\sl)} \, \Phi^{(w)}_{j;m,\mbar} ~,
} 
whose conformal weight and $J^3$ charges are 
\eqn[aass]{h\big[\Phihat^{(w)}_{j;m,\mbar}\big] = -\frac{j(j-1)}{k}-m w - \frac{k}{4}w^2
~~,~~~~
J_0^3\big[\Phihat^{(w)}_{j;m,\mbar}\big] = m+\frac{k}{2}w ~.}
The operators $\Phihat^{(w)}_{j;m,\mbar}$ are again superconformal primaries, but not current algebra primaries.

\vskip 1cm


\vskip 3cm

\bibliographystyle{JHEP}      

\bibliography{fivebranes}


\end{document}